\documentclass[12pt]{article}

\def\bi{\begin{itemize}}
\def\ei{\end{itemize}}
\def\bea{\begin{eqnarray}} 
\def\eea{\end{eqnarray}}
\def\be{\begin{equation}}
\def\ee{\end{equation}}
\def\line{\hbox to \hsize}    
\def\frac #1#2{{#1\over #2}}

\def\ad{ a^{\dagger}}

\def\ket #1{{\vert #1\rangle}}

\def\dbrak#1#2{{\langle#1\vert #2\rangle}}
\def\brak #1#2{{\langle#1, #2\rangle}}
\def\eval #1#2#3{{\langle#1\vert#2\vert#3\rangle}} 
\def\vev #1{{\langle #1\rangle}}
\def\1{\mbox{\bf I}}

\def\tanh{{\rm tanh\,}}

\def\bm#1{\mbox{\boldmath$#1$}} 

\newenvironment{Quote}
{\begin{list}{}{%
\setlength{\leftmargin}{10 pt}
\setlength{\rightmargin}{\leftmargin}}
\item[]}
{\end{list}}

     {\refstepcounter{exercisenumber} \begin{Quote}\small{\sl
     Exercise~\thechapter.\theexercisenumber\/}:}
{\end{Quote}}

{\begin{Quote}\small{\sl Example\/}:}
{\end{Quote}}

\def\levelonelist{
        \begin{list}{\mybulA}%
                        {
        \setlength{\topsep}{0pt}
        \setlength{\parsep}{0pt}
        \setlength{\partopsep}{0pt}
        \setlength{\itemsep}{0pt}
                        }
                }
\def\leveltwolist{
        \begin{list}{\mybulB}%
                        {
        \setlength{\topsep}{0pt}
        \setlength{\parsep}{0pt}
        \setlength{\partopsep}{0pt}
        \setlength{\itemsep}{0pt}
                        }
                }

\def\el{\end{list}}

\usepackage{amsfonts}
\usepackage{amssymb}
\usepackage{amsmath,mathtools}
\usepackage{centernot}
\usepackage{appendix}
\newcommand{\fsl}[1]{{\centernot{#1}}}
\usepackage{bm}

\begin{document}

\title{\bf Understanding  the chiral and parity anomalies without Feynman diagrams}
\vskip 6pt

\author{
Michael  Stone$^1$\\
\and   
JiYoung Kim$^2$\\
\and
Porter Howland$^3$ \\ \\
University of Illinois, Department of Physics\\ 1110 W. Green St.\\
Urbana, IL 61801 USA}   
\date{%
    $^1$m-stone5@illinois.edu\\%
    $^2$jkim623@illinois.edu\\%
    $^3$pbh2@illinois.edu \\[2ex]%
    \today
}
\maketitle

\begin{abstract} 
We review   the construction of the adiabatic expansion for Bose and Fermi  systems and show how  it may  be used to explore  the chiral  and parity anomalies for Dirac fermions without the need to compute  Feynman diagrams. 

\end{abstract}

\tableofcontents

\section{Introduction}  

Feynman diagrams are   the standard  tool for computing quantities in perturbative quantum field theory.  After one has learned the  rules for converting   diagram to integral,   and the often formidable techniques  necessary for  evaluating  the   resulting  integral, their power is such they provide  a magical black box into which one inserts a problem and extracts an answer.   What is often lost in the process is  a  picture  of what physics the mathematical machinery  is  capturing.  This is  true even at the level of  one-loop diagrams whose mathematics  can  output    non-obvious physical effects  such as  the ABJ chiral anomaly \cite{steinberger,adler,bell-jackiw},  the  parity-anomaly in odd space-time dimensions \cite{parity-anomaly,parity-anomaly2},  and the related current inflow  from higher dimensions that provides  the anomalous chiral charge  \cite {callan-harvey}.     

When using  one-loop diagrams to evaluate the vacuum expectation of  operators such as currents, charges, and energy fluxes we are basically exploring the physics of  systems whose hamiltonians are quadratic in annihilation and creation operators  and with coefficients that depend on  whatever  perturbations  are represented by the external legs on the diagram.  These perturbations can be electromagnetic fields  coupling  to the bilinear current  operator, or perhaps  the gravitational effects of curved space that couple to the energy-momentum tensor.     In such a case  the effects of slowly varying  external fields  can be captured  by a  gradient or derivative expansion of the one-loop effective action \cite{chan,aitchison,dunne-gradient}. In particular, if the external fields depend on only {\it one\/} space-time dimension a powerful tool is provided  by the adiabatic expansion of ordinary quantum mechanics, which does not require the full machinery of quantum field theory.   Indeed much of the work on field theory in curved space uses  exactly this tool \cite{birrell-davies,parker-toms}.

In this paper we will use versions of the  adiabatic expansion to exhibit   special cases of  the anomaly-related  effects mentioned above, and in doing so  hope to achieve some insights that are denied in the diagram derivations.   We   set the stage in section  \ref{SEC:QM}  by reviewing  how time-varying   parameters in a harmonic  oscillator  causes the ground-state to  evolve. We  relate this evolution  to vacuum squeezing in both Bose and Fermi systems and  
 establish the basic recurrence relations that allow us  to mechanically   compute the slow-squeeze adiabatic series  to arbitrary order. Then, in section \ref{SEC:Fields}, we   apply what we have learned   to field theory.   
 We use  the fermion version of the adiabatic series  in $1+1$ spacetime dimensions to 
 show how the standard spectral-flow picture of the chiral anomaly for massless fermions is affected  by the inclusion of a fermion mass. Similar methods are then used to  obtain the related parity anomaly in $1+2$ spacetime dimensions, and in $1+3$   dimensions to compute the  gradient expansion for the current induced by an external spatially constant electric field.   {\it En passant\/} we  obtain  the one-loop beta function for QED. Finally we extend the chiral anomaly results to four-dimensional spacetime.
  In the appendices  we  verify, when possible, the output  of our asymptotic expansions by comparing  them with  one-loop results obtained by other methods.  

We use units in which $\hbar=\epsilon_0=\mu_0=1$.

\section{Quantum Mechanics}
\label{SEC:QM}

\subsection{Time-dependent  harmonic  oscillators}
\label{SEC:bose}
 
The quantum harmonic oscillator with Hamiltonian 
\be
H_0= \frac 12 \hat p^2 + \frac 12 \Omega^2 \hat x^2
\label{EQ:harmonic-oscillator}
\ee
is considered in every introductory textbook --- not only because it is easily solved and  therefore  a pedagogically useful illustration, but also because  many real world systems are well approximated as   harmonic oscillators.  An  oscillator  driven by  a time-varying   linear term 
\be
 H(t) = \frac 12 \hat p^2 + \frac 12 \Omega^2 \hat x^2  + F(t) \hat x
\ee
 is also   useful and  straightforward  to solve (see appendix \ref{SEC:driven}) 
 but when it is the   oscillator  {\it frequency\/}  $\Omega$  that is allowed to depend on  time  the problem is    more challenging, and the physical effects  more exotic.
   A period of rapid frequency change  will  leave the oscillator in a  {\it squeezed state} --- a superposition of excited states that has many applications in quantum optics \cite{walls}, and   even in   gravitational wave detection \cite{ligo}.  We will see, however,  that   much can also be learned by  tracking   what happens {\it during\/} a slow  frequency change that leaves  little permanent excitation.

\subsection{Schr\"odinger-picture wavefunctions}
\label{SEC:wavefunction}

  The   wavefunction $\psi(x,t)$ of  a  variable-frequency harmonic oscillator   obeys the  time-dependent  Schr{\"o}dinger equation 
\be
i\frac{\partial \psi}{\partial t} = \left(-\frac 12 \frac{\partial^2}{\partial x^2}  +\frac 12 \Omega^2(t) x^2\right) \psi, \quad\Omega(t) \in {\mathbb R}.
\ee
This equation     has  a Gaussian solution \cite{husimi2}
\be
\psi(x,t)=  \chi^{-1/2}(t) \exp\left\{-\frac 12 \omega(t) x^2\right\}
\ee
provided that
 $\chi(t)$, $\omega(t)$ obey the  evolution equations 
\bea
{\dot\chi}/{\chi} &=&i\omega,\nonumber\\
\omega^2-i\dot \omega &=& \Omega^2(t).
\label{EQ:evolution}
\eea
Here the dot denotes a time derivative: $\dot \omega \equiv  \partial_t \omega$. 
With $\omega= \omega_R+i \omega_I$,
the   parameter-evolution equations imply that 
\bea
\partial_t \ln(|\chi |)&=& -\omega_I,\nonumber\\
\partial_t \ln(\omega_R)&=&+2 \omega_I,
\eea
and so ensure that   the normalization $\propto |\chi| \omega_R^{1/2}$ is preserved.

The   equations  (\ref{EQ:evolution}) together with the Riccati identity 
\be
-(-\partial_t-i\omega)(\partial_t-i\omega) = \partial^2_{tt} +(\omega^2 -i\dot \omega)
\ee
show that
\be
\frac{d^2\chi}{dt^2}+\Omega^2(t) \chi=0, \qquad 
\label{EQ:star}
\ee
where
\be
\chi(t)= \exp\left\{ i\int^t_{-\infty} \omega(\tau)\,d\tau\right\}.
\label{EQ:star-star}
\ee

We can rewrite ($\ref{EQ:star}$) as 
\be
\left(-\frac{d^2}{dt^2} +[\Omega_0^2-\Omega^2(t)]\right)\chi = \Omega^2_0 \chi,
\ee
and if $\Omega^2(t)\to \Omega^2_0$  as $t \to \pm \infty$, regard it as a  scattering problem with the frequency excursion away from $\Omega^2_0$ providing   the scattering potential. Consider  boundary conditions for which  the asymptotic solution is of the form
\be
\chi(t) \to
\begin{dcases} 
T e^{i\Omega_0 t},  & t\to -\infty ,\\
 e^{i\Omega_0 t}+ R e^{-i\Omega_0 t} ,& t \to +\infty,
 \end{dcases}
 \ee
with $|T|^2 =1-|R|^2$. For $t$ in the pre-excursion region the asymptotic  form for $\chi$ gives   
\be
\omega(t)=  -i\left(\frac {\dot \chi}{\chi}\right) {\to} \Omega_0
\ee
 so the ``transmission coefficient'' $T$ does not affect $\omega(t)$.
For $t$ in the post-excursion asymptotic region we have 
\be
\omega(t)= -i\left(\frac {\dot \chi}{\chi}\right) {\to} \Omega_0 \left(\frac{1-R\,e^{-2i \Omega_0t}}{1+R\,e^{-2i\Omega_0t}}\right).
\ee
The oscillations in $\omega(t)$ reveal that the gaussian wavefunction is    breathing in and out --- {\it i.e.\/} getting narrower and wider --- at frequency $2\Omega_0$.

Setting $y=0$ and $s=i\sqrt{R}e^{-i\Omega_0t} $ in Mehler's formula 
\be
\sum_{n=0}^\infty s^n \varphi_n(x)\varphi_n(y) = \frac
1{\sqrt{\pi (1-s^2)}} \exp\left\{\frac{4xys
-(x^2+y^2)(1+s^2)}{2(1-s^2)}\right\}, \quad  0\le |s|<1,
\ee
where
\be
\varphi_n(x)\equiv  \frac{1}{\sqrt{2^n n! \sqrt{\pi}}} H_n(x) e^{-x^2/2}
\ee
is the normalized $\Omega_0=1$  harmonic oscillator wavefunction, 
we find that 
\be
\frac{1}{(\pi\Omega_0)^{1/4}}\chi^{-1/2}(t) \exp\left\{-\frac 12 \omega(t) x^2\right\}\stackrel{t\to+ \infty}{=}\pi^{1/4}\sum_{n=0}^\infty e^{-i(n+1/2) \Omega_0 t} \varphi_n(0)(i\sqrt R)^n  \frac{\varphi_n(\sqrt{\Omega_0} x)}{(\Omega_0)^{1/4}}.
\ee
  Now 
$\varphi_n(0)$ vanishes if $n$ is odd, and 
\be
\pi^{1/4}\varphi_{2n}(0)= \frac{1}{\sqrt{4^n (2n)! } } \frac{(2n)!}{n!}(-1)^{n}.
\ee
Comparing with the wavefunction for $t\to -\infty$ we see that  the amplitude for being excited from the ground state to the $2n$-th eigenstate is 
\be
A_{2n}=\sqrt{T}  (R\,e^{-2i \Omega_0t})^n \frac{1}{\sqrt{4^n (2n)! } } \frac{(2n)!}{n!}.
\ee
As a check we may evaluate  
\be
\sum_{n=0}^\infty |A_{2n}|^2 = |T| \sum_{n=0}^\infty  |{\textstyle \frac 12} R|^{2n} \frac{(2n)!}{(n!)^2} =\frac{ |T|}{ \sqrt{1-|R|^2}}=1.
\ee
The probabilities of excitation therefore sum to unity as they should.

\subsection{Squeezed vacuum states}
\label{SEC:squeezed}
  We can appreciate  the formula for $A_{2n}$ by relating it to the  generalized coherent states  associated with the  non-compact group ${\rm SU}(1,1)\simeq {\rm Sp}(2,{\mathbb R})$ or, more accurately, with  its metaplectic double cover ${\rm MSp}(2,{\mathbb R})$.  In quantum optics these coherent states are   known as {\it squeezed vacuum states}.

Let $\hat a$, ${\hat a}^\dagger$ be  bosonic annihilation and creation operators with their usual commutation relation
$
[\hat a, {\hat a}^\dagger]=1,
$
and vacuum state $\ket{0}$ defined by $\hat a\ket{0}=0$.

A unitary  infinite-dimensional Fock-space representation of the   Lie  algebra $\mathfrak {su} (1,1)\simeq \mathfrak{sp}(2,{\mathbb R})$ is then generated by the quadratic operators $a^2$, ${ \ad}^2$ and ${\hat a}^\dagger \hat  a+\textstyle \frac12$ whose commutators are
\bea
[({\hat a}^\dagger)^2,   {\hat a}^2]&=& -4 ( {\hat a}^\dagger \hat a+\textstyle \frac12), \nonumber\\
{}[ ( {\hat a}^\dagger  \hat  a+{\textstyle \frac12}),   \hat a^2]&=&-2 \hat a^2,\nonumber\\
{}[( {\hat a}^\dagger  \hat a+{\textstyle \frac12}), ( {\hat a}^\dagger )^2]&=& +2  ({\hat a^\dagger})^2.
\eea
By exponentiating these generators we  construct  a unitary  {\it squeezing\/} operator \cite{stolerI}
\be
S(z)\stackrel{\rm def}{=} \exp\left\{{\textstyle \frac12}(z ({{\hat a}^\dagger})^2 -z^* \hat a^2)\right\}, 
\ee
which  implements the Bogoliubov-Valatin  transformation  
\bea
S^\dagger(z) \left[\begin{matrix} \hat a\cr {\hat a}^\dagger\end{matrix}\right]  S(z)&=& \left[\begin{matrix} \cosh|z| &  e^{i\theta}  \sinh |z|\cr 
 e^{-i\theta}  \sinh |z| & \cosh|z|\end{matrix}\right] \left[\begin{matrix} \hat a \cr {\hat a}^\dagger\end{matrix}\right].\
\eea
Here the angle $\theta$ is defined by $z=|z|e^{i\theta}$.

There is also a    faithful but non-unitary  representation of $\mathfrak{sp}(2,{\mathbb R})$ in terms of the two-by-two Pauli matrices in which
\bea
a^2&\mapsto& 2i\sigma_-,\nonumber\\
({\hat a }^\dagger)^2 &\mapsto& 2i\sigma_+,\nonumber\\
({\hat a}^\dagger \hat a+\textstyle \frac12)&\mapsto& \sigma_3.
\eea 
Because the representation is faithful, the  resulting group-element   map 
\be
 \exp\left\{{\textstyle \frac12}(z ({\hat a}^\dagger)^2 -z^* \hat a^2)\right\}\mapsto  \exp\left\{(iz\sigma_+  - iz^* \sigma_-)\right\}= \exp\left\{\left(\begin{matrix}0&iz\cr- iz^*&0\end{matrix} \right)\right\}.
\ee 
is an isomorphism. Consequently 
the   Gauss-Bruhat factorization  
\bea
\exp\left\{\left(\begin{matrix}0&iz\cr- iz^*&0\end{matrix} \right)\right\}&\equiv & \left(\begin{matrix}\cosh |z| & ie^{i\theta} \sinh |z| \cr -i e^{-i\theta}\sinh |z| &\cosh |z|\end{matrix}\right),\nonumber\\
&=&\left( \begin{matrix}1&ie^{i\theta} \tanh|z|\cr 0&1\end{matrix} \right)\left(\begin{matrix}1/\cosh|z|&0\cr 0 &\cosh |z| \end{matrix} \right)
\left(\begin{matrix}1&0\cr -ie^{-i\theta}\tanh |z|& 1\end{matrix} \right),\nonumber\\
&=&\exp\{ ie^{i\theta}  \tanh|z|\sigma_+\} \exp\{ - \ln(\cosh|z|) \sigma_3\} \exp\{-ie^{-i\theta}\tanh |z|\sigma_-\}\nonumber\\
\eea
of the two-by-two matrix 
establishes \cite{fisher} the normal-ordered factorization of   the  infinite-dimensional Fock-space operator
\bea
S(z)&=& \exp\left\{{\textstyle \frac12}(z ({\hat a}^\dagger)^2 -z^* \hat a^2)\right\}\nonumber\\
&=&\exp\left\{e^{i\theta}{\textstyle \frac12}\tanh |z|\, ({\hat a}^\dagger)^2\right\}\exp\left\{ -\ln\cosh |z| ({\hat a}^\dagger \hat a+{\textstyle \frac12})\right\} \exp\left\{-e^{-i\theta}{\textstyle \frac12}\tanh |z| \, \hat a^2\right\}.\nonumber\\
\eea

The  normal-ordering     shows that  
\bea
S(z) \ket{0}&=& \frac{1}{\sqrt{\cosh |z|} }\sum_{n=0}^\infty \frac{1}{n!} (e^{i\theta} {\textstyle \frac12}\tanh |z|)^n ({\hat a}^\dagger)^{2n}\ket{0} \nonumber\\
&=& \frac{1}{\sqrt{\cosh |z|}} \sum_{n=0}^\infty \frac{1}{n!} (e^{i\theta} {\textstyle \frac12} \tanh |z|)^n \sqrt{(2n)! }\ket{2n}\nonumber\\
&\stackrel{\rm def}{=}&  \sum_{n=0}^\infty A_{2n} \ket{2n}.
\eea
After identifying $\ket{n}$ with the $n$-th eigenstate of our oscillator and  $e^{i\theta}\tanh |z|$ with $Re^{-2i \Omega_0t}$  we  recognize the oscillator's  post-excursion excited state as a squeezed vacuum state,  and so understand the combinatoric origin of  
\bea
\sum_{n=0}^\infty |A_{2n}|^2 &=& \frac{1}{|\cosh |z||} \sum_{n=0}^\infty ({\textstyle \frac12} \tanh |z|)^{2n} \frac{(2n)!}{(n!)^2}\nonumber\\
&=& \frac{1}{|\cosh |z||}  \frac{1} {\sqrt{1- \tanh^2 |z|}}\nonumber\\
&=&1.
\eea

\subsection{Two-mode squeezed vacua} 
\label{SEC:two-squeezed}

Given  {\it two\/} frequency-$\Omega$ harmonic oscillators with ladder operators $\hat a$ and $\hat b$ we can similarly construct     an  operator
\bea
S_{2}(\xi)&\equiv& \exp\{\xi^*\hat{a}\hat{b}-\xi\hat{a}^\dagger\hat{b}^\dagger\} \nonumber\\&=& \exp\{-e^{i\theta}\tanh|\xi|\hat{a}^\dagger\hat{b}^\dagger\} \exp\{-\ln\cosh|\xi| ((\hat{a}^\dagger\hat{a}+{\textstyle \frac12})+(\hat{b}^\dagger\hat{b}+{\textstyle \frac12)})\} \exp\{e^{-i\theta}\tanh|\xi| \hat{a}\hat{b}\}\nonumber\\
\eea
that  creates  a  two-mode squeezed vacuum state
\bea
S_{2}(\xi)\ket{0}&=&\frac{1} {\cosh|\xi|} \exp\{-e^{-i\theta}\tanh|\xi|\hat{a}^\dagger\hat{b}^\dagger\} \ket{0}\nonumber\\
&=&  \frac{1} {\cosh|\xi|} \sum_{n=0}^\infty \frac 1{n!} (-e^{i\theta}\tanh|\xi|)^n ({\hat a}^\dagger{\hat b}^\dagger)^n \ket{0}\nonumber\\
&=& \frac{1} {\cosh|\xi|} \sum_{n=0}^\infty (-e^{i\theta}\tanh|\xi|)^n \ket{n,n}.
\eea
When we observe only the $\hat a$  mode,  the probability of being  in the  $n$-th  excited state  is 
\be
p_n= \frac{1}{\cosh^2 |z|} (\tanh^2 |z|)^n
\ee
which is  classical thermal Bose distribution with $e^{-\beta \Omega} = \tanh^2|z|$. If  we observe both the  $\hat a$ and $\hat b$  modes together we will find, however,  that they are  non-classically quantum entangled.

 We introduced the two-mode operator (\ref{EQ:2-mode-bose})  because we will later have cause  to refer to its    fermionic cousin. When    $\hat a$, $\hat a^\dagger$, $\hat b$, $\hat b^\dagger$  obey the fermion algebra 
\be
\{ \hat a, \hat a^\dagger\}= \{ \hat b, \hat b^\dagger\}=1, \quad \{ \hat a, \hat a\}=\{ \hat b, \hat b\}=\{ \hat a, \hat b\}= \{ \hat a, \hat b^\dagger\}=0,
\ee
we have $\hat a^2 =(\hat a^\dagger)^2=0$, so  there is no  fermion analogue of a  single-mode squeezing operator. 
We can, however,  still construct a  two-mode  operator 
\bea
U[z]&=&\exp\{z \hat a^\dagger \hat b^\dagger - z^* \hat b\hat a\}\nonumber\\
&=& \exp\{(e^{i\theta} \tan |z|) \hat a^\dagger\hat  b^\dagger\} \exp\{(\ln\cos |z|)[ (\hat a^\dagger \hat a+{\textstyle \frac 12})+( \hat b^\dagger \hat b+{\textstyle \frac 12 })]\}\exp\{(-e^{-i\theta} \tan |z|) \hat b \hat a\}\nonumber\\
\label{EQ:2-mode-bose}
\eea
which     also implements a  Bogoliubov-Valatin transformation 
\bea
U[z]\hat a U^\dagger[z] &=& (\cos |z|) \hat a - (e^{i\theta}\sin |z|) \hat b^\dagger, \nonumber\\
U[z]\hat b U^\dagger[z] &=&  (e^{i\theta}\sin |z|) \hat a^\dagger +(\cos |z|) \hat b.  
\label{EQ:fermi-bogoliubov}
\eea
The  right-hand-side of (\ref{EQ:fermi-bogoliubov}) is now a compact  ${\rm SU}(2)$ rotation rather than a non-compact ${\rm SU}(1,1)$ transformation. 

The   factored form  shows that  $U[z]$ acts on the vacuum to create a squeezed state  of the form 
\bea
U[z]\ket{0}
&=& \cos|z| \exp\{(e^{i\theta} \tan |z|) \hat a^\dagger \hat  b^\dagger\}\ket{0}\nonumber\\
&=&\cos|z|\{1+ (e^{i\theta} \tan |z|) \hat a^\dagger \hat  b^\dagger\}\ket{0}.
\label{EQ:2-mode-fermion}
\eea
If we  ascribe  an energy  $\epsilon $ to the $\hat a$ mode and define a real number $\beta$ so that 
\bea
\tan^2 |z|&=&e^{-\beta \epsilon},\nonumber\\
\sin^2 |z|&=& \frac{e^{-\beta \epsilon }}{1+e^{-\beta \epsilon }},\nonumber\\
\cos^2 |z|& =& \frac1{1+e^{-\beta \epsilon }},
\eea
  the probabilities of observing   the $\hat a$ mode as being unoccupied or occupied are respectively 
\be
p_0 = \frac1{1+e^{-\beta \epsilon}} , \quad p_1 = \frac{e^{-\beta \epsilon}}{1+e^{-\beta \epsilon}}.
\ee
This  is  again a thermal  distribution  \cite{khanna}, but now a Fermi one. As before the  $\hat a$ and $\hat b$ modes are quantum entangled.

\subsection{Squeezing in the Heisenberg picture}
\label{SEC:heisenberg}

The  rhythmic  in-and-out out breathing of the time-dependent Shr\"odinger wavefunction   gives a concrete physical picture  of the effect  of squeezing  on  an    oscillator ground state.  For applications to field theories, however,   it  is  more  convenient   to work in  the  
 Heisenberg-picture    language where  the states  do not evolve  but  instead    the Hermitian position and momentum  operators    $\hat x$, $\hat p$    depend on time and obey both their classical  equations of motion and the  quantum equal-time commutation relation
\be
[\hat x(t), \hat p(t)]=  {i}.
\ee
For the   variable-frequency  harmonic  oscillator with Hamiltonian 
\be
H= \frac 1{2}\hat p^2+\frac 1 2 \Omega^2(t) \hat x^2
\ee
the classical equation of motion is 
\be
\frac{d^2\hat x }{dt^2}+\Omega^2(t) \hat x=0,
\ee
and $\hat p(t)\equiv \dot {\hat x}(t)$ so the  commutation relation is $[\hat x(t),  \dot {\hat x}(t)]=i$.

The equation of motion is linear, so we  can expand the Hermitian operator  
$\hat x(t)$ as a sum
\be
\hat x(t)= f(t) \hat a +f^*(t) \hat a^\dagger,
\ee
where the  constant coefficients $\hat a$ and $\hat a^\dagger$ are   operators and the complex-valued  $c$-number function $f$ obeys  
\be
\ddot f+\Omega^2(t)f=0.
\ee
The condition $[\hat x(t), \dot {\hat x}(t)]= {i}$ requires the   coefficients   $\hat a$, $\hat a^\dagger$ to obey
\be
\brak{f}{f}[\hat a,\hat a^\dagger]=1, \quad \hbox{where} \quad  \brak{f}{g}\stackrel{\rm def}{=} i( f^*\partial_t g- (\partial_t f^*)g).
\ee
Being proportional to the   Wronskian,  the non-positive-definite ``inner product'' $\brak{f}{g}$ is independent of $t$, so there is no contradiction with $\hat a$ and $\hat a^\dagger$ being constants.

For   constant $\Omega$  the   appropriate choice for making $[\hat a,\hat a^\dagger]=1$ is  to take $f$ as  the {\it positive-frequency\/} solution 
\be
f(t)=\sqrt{\frac 1{2\Omega}}e^{- i\Omega t}.
\ee
With this choice  
\be
H=
\Omega(\hat a^\dagger \hat a +{\textstyle \frac 12} ),
\ee 
so the ground state $\ket{0}$ obeys $\hat a\ket{0}=0$
and  we have the  Heisenberg-picture expansions
\bea
\hat x(t)&=& \sqrt{\frac 1{2\Omega}} (\hat a^\dagger e^{i\Omega t } +\hat ae^{-i\Omega t}),\nonumber\\
\hat p(t)&=& i  \sqrt{\frac{\Omega }{2}} (\hat a^\dagger e^{i\Omega t } -\hat ae^{-i\Omega t }),
\eea
\bea
\hat a\,e^{-i\Omega t}&=& \frac{1}{\sqrt{ 2}}\left(\sqrt{\Omega}\, \hat x(t)+\frac{i}  {\sqrt{\Omega} }\hat p(t)\right),\nonumber\\
\hat a^\dagger e^{i\Omega t}&=&\frac{1}{\sqrt {2}}\left(\sqrt{\Omega}\, \hat x(t) -\frac{i}  {\sqrt{\Omega} }\hat p(t)\right).
\eea

Now consider a frequency excursion with $\Omega_{\rm in}$, $\Omega_{\rm out}$ as the initial and final asymptotic values of $\Omega(t)$.  If we start in the initial Heisenberg-picture ground  state $\ket{0}_{\rm in}$   neither the state nor  the     $a_{\rm in}$ and $a^\dagger_{\rm in}$ coefficients change 
but a c-number solution that starts off as 
\be
f(t)= \sqrt{\frac {1}{2\Omega_{\rm in }}} e^{-i\Omega_{\rm in} t }
\ee
in the distant past will evolve to 
\be
\alpha \sqrt{\frac {1}{2\Omega_{\rm out  }}} e^{-i\Omega_{\rm out} t }+\beta \sqrt{\frac {1}{2\Omega_{\rm out }} }e^{i\Omega_{\rm out} t }
\ee
after  the frequency has ceased to change. It is natural to define new expansion coefficients  $\hat a_{\rm out}$ and  $\hat a^\dagger _{\rm out}$ by writing
\bea
\hat x(t) &=& \hat a_{\rm in} \left(\alpha \sqrt{\frac {1}{2\Omega_{\rm out  }}} e^{-i\Omega_{\rm out} t }+\beta \sqrt{\frac {1}{2\Omega_{\rm out }}}e^{i\Omega_{\rm out} t }\right)\nonumber\\
&&\quad + \hat a^\dagger_{\rm in} \left(\alpha^* \sqrt{\frac {1}{2\Omega_{\rm out  }} }e^{i\Omega_{\rm out} t }+\beta^* \sqrt{\frac {1}{2\Omega_{\rm out }}} e^{-i\Omega_{\rm out} t }\right)\nonumber\\
&\stackrel{\rm def}=&\hat a_{\rm out} \sqrt{\frac {1}{2\Omega_{\rm out  }} }e^{-i\Omega_{\rm out} t }+\hat a^\dagger_{\rm out} \sqrt{\frac {1}{2\Omega_{\rm out  }} }e^{i\Omega_{\rm out} t }.\
\eea
Comparison of the last two lines shows that  
\bea
\left[ \begin{matrix} \hat a_{\rm out}\cr \hat a^\dagger _{\rm out}\end{matrix}\right] &=&
 \left[ \begin{matrix} \alpha \,& \,\beta^* \cr \beta \,&\,\alpha^* \end{matrix}\right] \left[ \begin{matrix} \hat a_{\rm in}\cr \hat a^\dagger _{\rm in }\end{matrix}\right]. 
 \eea
 The commutation relation for the ``out" operators require that 
  $|\alpha^2|-|\beta|^2=1$, which holds true because the Wronskian is constant. Using this  we can solve for inverse transformation  
 \bea
 \left[ \begin{matrix} \hat a_{\rm in}\cr \hat a^\dagger _{\rm in }\end{matrix}\right] &=&
 \left[ \begin{matrix} \alpha^* & -\beta^* \cr- \beta &\alpha \end{matrix}\right] \left[ \begin{matrix} \hat a_{\rm out}\cr \hat a^\dagger _{\rm out }\end{matrix}\right]. 
 \eea
 
The initial state   $\ket{0}_{\rm in }$ now appears as   a  squeezed version
\be
\ket{0}_{\rm in}=  \exp\left\{{\textstyle \frac12}(z (\hat a^\dagger_{\rm out})^2 -z^* \hat a_{\rm out}^2)\right\} \ket{0}_{\rm out},
\ee
of  the $\Omega_{\rm out}$ ground state    defined by $\hat a_{\rm out}\ket{0}_{\rm out}=0$. Here
\bea
\alpha &=& \cosh|z|,\nonumber\\
\beta^*&=& e^{i\theta} \sinh |z|.
\eea

That the oscillator is not in the state $\ket{0}_{\rm out}$  manifests itself  through the computation of the energy expectation
\bea
_{\rm in}\eval{0}{{\textstyle \frac 12}  \dot {\hat x}^2+ {\textstyle \frac 12} \Omega^2_{\rm out} \hat x^2}{0}_{\rm in}&=& {\textstyle\frac 12} \Omega_{\rm out} (  |\alpha|^2 +|\beta|^2 )\nonumber\\
&=&  \Omega_{\rm out} \left(|\beta|^2+ {\textstyle \frac 12}\right)\nonumber\\
&> & {\textstyle \frac 12}  \Omega_{\rm out} ,
\eea
and matrix elements such as  
\bea
_{\rm in}\eval{0}{\hat x^2}{0}_{\rm in}&=& \left|\alpha \sqrt{\frac {1}{2\Omega_{\rm out  }}} e^{-i\Omega_{\rm out} t }+\beta \sqrt{\frac {1}{2\Omega_{\rm out }} }e^{i\Omega_{\rm out} t }\right|^2\nonumber\\
&=&\frac {1}{2\Omega_{\rm out} }  \left\{ |\alpha|^2 +|\beta|^2 + (\alpha^*\beta e_+^2+  \alpha\beta^* e_-^2)\right\},
\label{EQ:matrix-element1}
\eea
where
\be
e_\pm(t) = \exp\left\{\pm i \Omega_{\rm out}t\right\}.
\label{EQ:epmdef}
\ee
Similarly
\bea
_{\rm in}\eval{0}{\dot {\hat x}^2}{0}_{\rm in}&=& \frac { \Omega_{\rm out  }}{2}  \left\{ |\alpha|^2 +|\beta|^2 -(\alpha^*\beta e_+^2+  \alpha\beta^* e_-^2)\right\},\nonumber\\
_{\rm in}\eval{0}{{\textstyle \frac 12}(\hat x\dot {\hat x}+\dot {\hat x} \hat x)}{0}_{\rm in}&=&\frac{i}{2}
(\alpha^*\beta e_+^2-  \alpha\beta^* e_-^2).
\label{EQ:matrix-element2}
\eea
The expectation values therefore show the same $2\Omega_{\rm out}$ pulsations as in the wavefunction description and we can identify  the transmission and reflection coefficient from section \ref {SEC:wavefunction} as $T=1/ \alpha $ and $R=\beta/\alpha$.

At intermediate times one  can seek  a solution of the form 
\be
f(t)= \alpha(t) \frac{1}{\sqrt{2\Omega(t)}} e_-(t) + \beta(t)  \frac{1}{\sqrt{2\Omega(t)}} e_+(t)
\label{EQ:intermediate}
\ee
where 
\be 
e_\pm(t) \stackrel{\rm def}{=} \exp\left\{\pm i \int^t \Omega(t')dt'\right\},
\label{EQ:epmdef2}
\ee
is a generalization of (\ref{EQ:epmdef}) to admit a variable frequency. Given such a solution we are invariably  tempted to  interpret the  quantity 
$|\beta(t)|^2$ as the average occupation number of the excited states above the  ground state of $H(t)$.  We may, however,  swap  terms  $\propto e^2_\pm $  between  $\alpha(t)$ and $\beta(t)$   and as a result the decomposition of $f(t)$ into positive and negative  frequency terms is not unique. This  non-uniqueness    makes  any  physical interpretation of $|\beta|^2$  unclear, and complicates  any  interpretation of $\ket{0}_{\rm in}$ as a squeezed ground state of  $H(t)$   \cite{dunne}.  
What {\it is\/}  well defined at all times are the  equal-time expectation values $_{\rm in}\eval{0}{\ldots }{0}_{\rm in}$ of  functions of the  Heisenberg-picture operators. These can  be extracted from the well-defined $f(t)$ alone,  and hence from the $\alpha(t)$ and $\beta(t)$ coefficients despite their individual ambiguity.

\subsection{Hyperbolic Bloch equations}
\label{SEC:hyperbolic}

One of many ways  of defining  $\alpha$ and $\beta$ coefficients during  the evolution of the system as $\Omega(t)$ varies 
is that of  Zeldovich and Starobinskii (ZS) \cite{ZS}
who    use Lagrange's method of variation of parameters to solve   
\be
\ddot \chi+\Omega^2(t) \chi=0. 
\label{EQ:star2}
\ee
ZS start by assuming the  ambiguous form (\ref{EQ:intermediate})
\be
\chi(t)= \alpha(t) \frac{1}{\sqrt{2\Omega}} e_-(t) + \beta(t)  \frac{1}{\sqrt{2\Omega}} e_+(t), \quad e_\pm(t) \stackrel{\rm def}{=} \exp\left\{\pm i \int^t \Omega(t')dt'\right\},
\label{EQ:ZS1}
\ee
but  follow Lagrange by demanding that 
\be
\dot \chi(t) = -i\Omega \left(\alpha(t) \frac{1}{\sqrt{2\Omega}} e_- - \beta(t)  \frac{1}{\sqrt{2\Omega}} e_+\right).
\label{EQ:ZS2}
\ee
This   expression is what we would obtain from differentiating $\chi(t)$ while  taking   $\alpha$, $\beta$, and $\Omega$  to be constants ---  but as  these quantities  vary with $t$  the demand   imposes the condition 
\be
0=\left(-\frac 12 \frac{\dot \Omega}{\Omega}\alpha + \dot \alpha\right) e_-+  \left(-\frac 12 \frac{\dot \Omega}{\Omega}\beta + \dot \beta\right) e_+\label{EQ:condition1}.
\ee 
This condition   serves to   uniquely specify   $\alpha(t)$ and $\beta(t)$ and hence to disambiguate  the decomposition of $\chi(t)$ into positive and negative frequency modes. 
 In particular, the  time independence of the Wronskian of $\chi$ and $\chi^*$ constructed using   (\ref{EQ:ZS1}) and (\ref{EQ:ZS2})    shows that  $|\alpha|^2-|\beta|^2=1$ at all times.

Inserting  the $\dot \chi(t)$ defined by (\ref{EQ:ZS2}) into (\ref{EQ:star2})  gives 
\be
0=\left(\frac 12 \frac{\dot \Omega}{\Omega}\alpha + \dot \alpha\right) e_--  \left(\frac 12 \frac{\dot \Omega}{\Omega}\beta + \dot \beta\right) e_+.
\label{EQ:condition2}
\ee
Adding and subtracting the two conditions  (\ref{EQ:condition1}) and (\ref{EQ:condition2})  we find
\bea
\dot \alpha &=&\frac 12 \frac{\dot \Omega}{\Omega} \beta e_+^2,\nonumber\\
\dot \beta  &=&\frac 12 \frac{\dot \Omega}{\Omega} \alpha e_-^2.
\label{EQ:dot-alpha}
\eea
It is difficult  to get a sense of what the equations (\ref{EQ:dot-alpha}) imply for the  evolution of $\alpha$ and $\beta$ because of the rapidly varying phase factors $e_\pm^2$. To deal with this  ZS
introduce the real quantities 
\bea
\sigma(t)&=&|\beta|^2,\nonumber\\
\tau(t)&=&i(\alpha\beta^*e_-^2 - \alpha^*\beta e_+^2), \nonumber\\
\upsilon(t)&=&(\alpha\beta^*e_-^2 +\alpha^*\beta e_+^2),  
\label{EQ:hyperbolic-hopf}
\eea
which   have already appeared in equations (\ref{EQ:matrix-element1}, \ref{EQ:matrix-element2}). These combinations obey 
$$
(1+2 \sigma)^2 - \tau^2-\upsilon^2=1,
$$
which  is  the   equation for a hyperboloid of two sheets. 
Indeed the manifold of squeezed vacuum states possesses an inherent   hyperbolic 
geometry arising  from it being a coset $K = {\rm Sp}(2,R)/{\rm U}(1)$ which can be identified as the  upper  sheet of a  two-dimensional hyperboloid embedded in 2+1 dimensional Minkowski space -- a classic model for Bolyai-Lobachevskii space. The  map (\ref{EQ:hyperbolic-hopf}) taking  $\alpha, \beta$ to  points on the coset is a hyperbolic version of the  ${\rm SU}(2)\to S^3$ Hopf map.

The   advantage of the  quantities  $\sigma$, $\tau$, $\upsilon$ is that when   $\alpha\approx 1$ and $\beta$ is small  the rapidly varying phases $e_\pm^2(t)$ almost cancel their rapid phase variation  in $\beta(t)$ and $\beta^*(t)$    and  allow   $\sigma$, $\tau$, and $\upsilon$    be slowly varying.  

Using the equations for $\dot \alpha$, $\dot \beta$ shows that 
\bea
\dot \sigma&=&\frac 12 \left(\frac{\dot \Omega}{\Omega}\right) \upsilon,\nonumber\\
\dot \upsilon&=&\left( \frac{\dot \Omega}{\Omega}\right)(1+2\sigma) -2\Omega \tau,\nonumber\\
\dot \tau&=& 2\Omega \upsilon.
\label{EQ:elliptic-bloch}
\eea
This set of three equations is a hyperbolic analogue of the Bloch equations describing the interaction of a spin  with a time dependent  magnetic  field.
The initial conditions $\alpha=1$, $\beta=0$ corresponds to  conditions on $\sigma,\tau,\upsilon$ that they are all zero in the distant past.

\subsection{Adiabatic expansion}
\label{SEC:adiabatic}

Much of the effort in investigating the squeezed  oscillator has focused  on the permanent excitation of the system after  the frequency excursion.  Such an  excitation is most efficiently achieved  by a period in which   $\Omega(t)$  oscillates   at a frequency near $2\Omega_0$  (parametric resonance) or by    relatively violent  changes in the frequency such as occur in Landau-Zener tunneling  \cite{zener}.  For  applications and  a review of  techniques  see \cite{vlasov,dunne-review}.
In the rest of this paper, however,  we will focus on the behavior of systems  {\it during\/}  a relatively slow {\it adiabatic\/}   frequency excursion in which little or no  permanent excitation occurs.  

There are  a number of methods for obtaining a systematic  adiabatic expansion of the time evolution of $\chi(t)$, and hence for quantities such as $\sigma(t)$.   A textbook  route 
\cite{birrell-davies,parker-toms}  starts from a single-exponential  WKB-like solution
\be
\chi(t) =\frac  1{\sqrt{2W(t)}}\exp\left\{-i \int^t W(\tau) d\tau\right\},
\ee
and generates a series expansion for $W(t)$. We will, however, continue with the ``two-exponential" method of \cite{ZS}  as it is  computationally simpler. 

 We   rearrange  the evolution  equations (\ref{EQ:elliptic-bloch})  as
\bea
\sigma&=& \frac 12 \int_0^t (\dot \Omega/\Omega) \upsilon dt,\nonumber\\
\tau&=& \frac{(\dot \Omega/\Omega)(1+2 \sigma) -\dot \upsilon}{2\Omega},\nonumber\\
\upsilon&=& \frac{\dot \tau}{2\Omega},
\eea
and  
 expand   in inverse powers of $\Omega(t)$ as 
\bea
\sigma&=& \sigma_{\{2\}}+ \sigma_{\{4\}}+\ldots\nonumber\\
 \tau&=& \tau_{\{1\}}+ \tau_{\{3\}}+\ldots\nonumber\\
 \upsilon &=& \upsilon_{\{2\}}+ \upsilon_{\{4\}}+\ldots.
\eea
Regarding $(\dot \Omega/\Omega)$ as being of $O[\Omega^0]$, we  obtain   recursion relations
\bea
\sigma_{\{n\}}&=& \frac 12 \int_0^t (\dot \Omega/\Omega) \upsilon_{\{n\}}dt,\nonumber\\
\upsilon_{\{n\}}&=& \dot\tau_{\{n-1\}}/2\Omega,\nonumber\\
\tau_{\{n+1\}}&=& \frac{2(\dot \Omega/\Omega) \sigma_{\{n\}}-\dot \upsilon_{\{n\}}}{2\Omega},
\eea
with starting condition $\tau_{\{1\}}= \dot\Omega/{2\Omega^2}$.

By hand we find, for example,
\be
\sigma_{\{2\}} = \int_0^t \frac{\dot \Omega}{4\Omega^2} \frac{d}{dt} \!\!\left[\frac{\dot \Omega}{2\Omega^2}\right] dt= \frac 1{16}\frac{ \dot \Omega^2}{\Omega^4}.
\ee
Using Mathematica to automate the labour, we  can compute higher order terms
\be
\sigma_{\{4\}}=\frac{\ddot \Omega^2}{64 \Omega^6}-\frac{45\dot  \Omega^4}{256 \Omega^8}-\frac{\Omega ^{(3)}\dot \Omega}{32 \Omega^6}+\frac{5 \dot \Omega^2
  \ddot  \Omega}{32 \Omega^7},
\ee
\bea
\sigma_{\{6\}}&=&\frac{(\Omega^{(3)})^2}{256 \Omega^8}+\frac{7 \ddot \Omega^3}{128 \Omega^9}+\frac{4725\dot  \Omega^6}{2048 \Omega^{12}}+\frac{\Omega ^{(5)}\dot \Omega}{128 \Omega^8}-\frac{\Omega^{(4)} \ddot\Omega}{128 \Omega^8}-\frac{7
   \Omega ^{(4)} \dot \Omega^2}{64 \Omega^9}\nonumber\\
  &&\quad  +\frac{217 \Omega ^{(3)}\dot  \Omega
   (x)^3}{256 \Omega^{10}}-\frac{945\dot  \Omega^4 \ddot\Omega}{256 \Omega^{11}}+\frac{441\dot  \Omega^2 \ddot \Omega^2}{512 \Omega^{10}}-\frac{21 \Omega
   ^{(3)} \dot \Omega \ddot \Omega}{128 \Omega^9},
\eea
and so on.

All terms in these expressions contain derivatives of $ \Omega(t)$. Consequently  they revert to being  zero when $ \Omega$ ceases to change. In particular, despite the appearance  of the factor  $\beta(t) \exp\{+ i \int^t \Omega(t')dt'\}$ in $\chi(t)$, the  recursion process  does  not generate the  ``reflected'' wave that indicates a permanent excitation of the oscillator. Such a   persistent  excitation is a non-perturbative effect  \cite{berry} so the expansion  is,  at best, an asymptotic series. Furthermore  because of the rapid $e_-(t)^2$ phase variation in  $\beta(t)$  the factor $\beta(t) \exp\{+ i \int^t \Omega(t')dt'\}$ is close in time evolution to $\alpha(t)\exp\{- i \int^t \Omega(t')dt'\}$. This  means   the ``two-exponential" expansion used by Zeldovich and Starobinskii \cite{ZS} is  consistent with the ``one-exponential" WKB expansion of \cite{birrell-davies,parker-toms} and nicely      illustrates the ambiguity in the notion of positive or negative frequency. 

The ambiguity means   our  asymptotic expressions for $\sigma=|\beta|^2$ are not in themselves physically meaningful ---  they depend on the choice we made when we imposed Lagrange's condition (\ref{EQ:condition1}).  However, when used as ingredients for computing quantities such as (\ref{EQ:matrix-element1}),(\ref{EQ:matrix-element2}) and the expectation values of the currents  that will appear in section \ref{SEC:Dirac}, the results {\it are\/} independent of this choice. 

We have been rather vague as to what exactly  is the small parameter in the adiabatic series. We  can make it explicit  by  replacing  $\Omega(t)$ by $\Omega(\varepsilon t)$ where $\varepsilon$ is to be assumed  small.  Then  $\sigma_{\{2n\}}$ is replaced by  $\varepsilon^{2n} \sigma_{\{2n\}}$ and  $\varepsilon$ becomes  the  small parameter defining  the resulting asymptotic expansion. We prefer, though,  to simply count the number of derivatives of $\Omega(t)$ in our expressions as this amounts to the same thing.

\section{Field Theory}
\label{SEC:Fields}

\subsection{Dirac fermions}
\label{SEC:Dirac}

Now we turn to the effect of changing   oscillator  frequencies in fermion systems.
In particular we will apply the adiabatic expansion to Quantum electrodynamics  (QED), the theory   of a spin-1/2 Dirac Fermi field coupled to   electromagnetism.  

With the  space-time  signature $(+,-,-,-)$   the four-vector  gauge field decomposes into   time   and space parts as   $A^\mu=(\phi, {\bf A})$ and   $A_\mu = (\phi, -{\bf A})$, where ${\bf A}=(A_x,A_y,A_z) $ is the usual three-vector potential in terms of which  ${\bf B}= \nabla\times {\bf A}$ and  ${\bf E}= - \nabla \phi - \partial_t {\bf A}$.  

On its own, the Maxwell gauge field is described by the action functional
\be
S_{\rm Maxwell}[A] =- \frac 1{4 e^2} \int F_{\mu\nu}F^{\mu\nu} d^d x= \frac 1{2 e^2} \int ({\bf E}^2- {\bf B}^2) d^d x.
\ee  
After being integrated-out in a path-integral formalism,  the Fermi field adds to $S_{\rm Maxwell}[A]$  the fermionic {\it effective action\/}    
\be
S_F[A]= - i \ln {\rm Det}({\fsl D}[A]+m), 
\ee
where ${\fsl D}[A]= i\gamma^\mu(\partial_\mu+i A_\mu) $. 

The  effective action is a sophisticated  object, being shorthand for   an infinite  sum  of one-loop Feynman diagrams with an arbitrary number of $\gamma^\mu A_\mu$ vertices. Even if we take the $A_\mu$ to be a non-fluctuating external field  $S_F[A]$ captures  much  interesting physics. For a static   magnetic field $S_F[A]$ is real number equal to minus the energy of the of the electrons in the field --- a relativistic analogue of the  De Haas-Van Alphen effect.  For a static    electric field $2 S_F[A]$ gains an imaginary part  that gives the rate of electron-positron pairs created per unit volume. For non-constant fields $S_F$ captures the 1-loop renormalization effects and the non-linear effect of  scattering of light by light. With   an extra $\gamma^5$ vertex inserted, we also uncover   the ABJ chiral anomaly \cite{steinberger,adler,bell-jackiw}.  

It is  not easy to compute $S_F[A]$ for  general space-time dependent $A_\mu$,  but with some effort one can compute  the first few terms in a systematic  expansion in powers of derivatives of $A_\mu$ \cite{chan,aitchison,dunne-gradient}.  When, however, we restrict ourselves to $A_\mu$ fields that depend on only {\it one\/} space-time coordinate   we can  --- after some small changes to take into account that we are dealing with fermions rather than bosons ---    exploit  the adiabatic expansion methods from section \ref{SEC:adiabatic} and   calculate many terms with relative ease.   
As an illustration we will consider the specific case of a spatially uniform, but time dependent,  electric field ${\bf E}(t)$  in  one, two and three space dimensions.

\subsection{One space dimension: chiral  anomaly}
\label{SEC:1-d-chiral}

To describe  our uniform electric field we will use a   gauge in which $A_0=0$ and  ${\bf E}=-\partial_t  {\bf A}$.
In  the one-dimensional case   the   Hamiltonian form of the Dirac equation  in this  field  becomes   
\be 
i\partial_t  \hat \psi= H(t) \hat \psi
\ee
where
\be
H(t) = -i\sigma_3(\partial_x -iA_x(t)) + \sigma_1 m
\ee
is a  differential operator involving the two-by-two Pauli matrices.

The 
field  equations for the Heisenberg-picture  operators $\hat \psi$ and $\hat \psi^\dagger$ are  
\bea
i\partial_t \hat \psi &=& - i\sigma_3(\partial_x -iA_x)\hat \psi+\sigma_1 m\hat \psi, \nonumber\\
-i\partial_t \hat \psi^\dagger &=&+i (\partial_x  +iA_x)\hat\psi^\dagger\sigma_3 +m \hat\psi^\dagger \sigma_1.
\label{EQ:dirac-field-equations}
\eea
The  equations (\ref{EQ:dirac-field-equations}) are linear so, as for the harmonic oscillator, 
the  operators  can be expanded  linear combinations 
\be
\hat \psi(x,t) = \int_{-\infty}^\infty \frac{dp}{2\pi} \Big(\hat a_p{\bm  \psi}_+(p,x,t)+ \hat b^\dagger_{-p} {\bm \psi}_-(p,x,t)\Big) 
\ee
of  two linearly-independent c-number solutions 
\be
{\bm \psi}_\pm (x,t)= \left[\begin{matrix} u_\pm (x,t)\cr v_\pm(x,t)\end{matrix}\right]
\ee
of the 
  equation of motion
\be
i\partial_t \left[\begin{matrix} u\cr v\end{matrix}\right]=  \left[\begin{matrix} -i(\partial_x -iA_x(t))  &m \cr m &+i(\partial_x-iA_x(t)) \end{matrix}\right]\left[\begin{matrix} u\cr v\end{matrix}\right].
\label{EQ:diracEOM}
\ee
As we are considering only spatially uniform systems  we have ${\bm \psi}_\pm(p,x,t)= {\bm \psi}_\pm(p,t) e^{ipx}$. 
In the Heisenberg picture the  operator-valued expansion coefficients $\hat a_p$ and $\hat b_{-p}^\dagger$ are time-independent   and obey  the standard  Fermi anti-commutation relations $\{\hat a_p, \hat a^\dagger_q\}= 2\pi \delta (p-q)$, {\it etc\/}.

 By formal manipulation of the field equations (\ref{EQ:dirac-field-equations}) we obtain  the  particle-number current   conservation equation 
\be
\partial_t (\hat \psi^\dagger \hat \psi) +\partial_x (\hat \psi^\dagger\sigma_3\hat \psi) =0, 
\ee
and, more interestingly,  the  
 chiral   current (non)-conservation equation in the form 
\be
\partial_t (\hat \psi^\dagger \sigma_3\hat \psi) +\partial_x (\hat \psi^\dagger\hat \psi) \stackrel{\rm ?}{=} 2m(\hat  \psi^\dagger \sigma_2 \hat \psi).
\ee
The ``?''  is there because, while  last equation is valid for  any {\it c-number\/}  solution ${\bm \psi}(x,t)$ of the field equations,  the actual 
equation obeyed by the {\it operator-valued\/}  chiral charge $\hat \psi^\dagger \sigma_3\hat \psi$  should be 
 \be
\partial_t (\hat \psi^\dagger \sigma_3\hat \psi) +\partial_x (\hat \psi^\dagger\hat \psi) = 2m(\hat  \psi^\dagger \sigma_2 \hat \psi) + \frac 1{\pi}E(t),
\label{EQ:1-d-chiral-anomaly}
\ee
where  extra term $E(t)/\pi$ is the 1+1 dimensional version of the chiral anomaly \cite{steinberger,adler,bell-jackiw}.

To understand the source of the $E/\pi$ term  it helps to visualize the initial many-body ground state  as a filled Dirac sea in which all   negative-energy states are   occupied and all positive-energy states vacant\footnote{The Dirac sea is rather  out of fashion, but the principal objection to it --- the associated infinite vacuum charge --- is rendered moot by the observation that each generation in the standard model has a charge-neutral sea: there are three (for colour) up-type quarks each with $q=+2/3$, three down-type quarks  with $q=-1/3$ and one electron-like lepton with $q=-1$. These charges  sum to zero.}. The antiparticle creation operator  $\hat b^\dagger_{-p}$  is  then to be  thought of as  an operator annihilating a  negative-energy positively charged particle that was occupying a momentum $+p$ mode and as a result   creating  a negatively-charged particle  excitation with  {\it positive\/}  energy  $\sqrt{p^2+m^2}$  
and   momentum $-p$.  

When $m=0$ and the field $E$  time independent  this view  of the ground state allows a simple    physical picture \cite{kiskis,ambjorn,peskin} of the how  the anomaly arises.
The   Hamiltonian  for the massless two component spinor is  now diagonal and the equation of motion  
\be
i\partial_t \left[\begin{matrix} u\cr v\end{matrix}\right] = \left[\begin{matrix} p+Et & 0\cr0  &-(p+Et) \end{matrix}\right]\left[\begin{matrix} u\cr v\end{matrix}\right]
\ee
can be interpreted  as stating that    the energy of a right-going mode which possessed  momentum $p$ at $t=0$ becomes  $p+Et$. Similarly the   energy of a left-going mode becomes  $p-Et$.  For positive $E$ there  is therefore   a steady flow  
 of   occupied right-going states into the positive energy continuum  and  an equal flow of empty left-going states  down into the negative continuum. The latter    leaves behind a growing number of holes and so,  compared to the ground state  at $t=0$, a negative left-going particle number. If  the ``volume'' of the system is $L$  the  density of  momentum modes  is  $dn/dp= L/2\pi$ so    the  chiral particle number    
 $ 
 n_R-n_L 
 $
 is changing according to  
\be
\frac 1 L \frac{d(n_R-n_L)}{dt}= 2 \times \frac{E}{2\pi}= \frac E \pi.
\ee
Taking the   chiral charge density operator to be    $\hat \psi^\dagger_R\hat \psi_R -  \hat \psi^\dagger_L\hat \psi_L= \psi^\dagger \sigma_3\hat \psi$ 
we have a $c$-number version
\be
 \frac{d \vev{\hat \psi^\dagger \sigma_3\hat \psi}}{dt}= \frac E{\pi}
 \ee 
 of Eq.\ ($\ref{EQ:1-d-chiral-anomaly}$).

 This  simple picture requires massless fermions. A mass term thwarts  the spectral flow: the negative energy right-going single-particle modes still  rise adiabatically  in energy from deep in the  sea but in the neighbourhood of   the  mass gap  they  mix with, and mutate into,   left-going modes which descend again into the depths. When  we ignore the possibility of Zener tunneling across the gap, the negative energy states remain filled and the positive energy ones empty,  so  $d(n_R-n_L)/{dt}$  is  zero. The  corresponding $c$-number version of the anomaly equation should   therefore  reduce to  
\be
2m \vev{\hat \psi^\dagger \sigma_2 \hat \psi} + \frac E{\pi}=0.
\ee
We explore how this comes about, and how the equation is modified when   the electric field $E$  depends on time.

\subsection{Adiabatic expansion for Dirac fields}
\label{SEC:Two-level-adiabatic}

Despite much work on the adiabatic expansion for  scalar Bose fields, the  construction of    systematic adiabatic expansions for the  time-dependent Dirac equation appears to be  relatively recent.   Landete, Navarro-Salas and Torrenti  \cite{landete}    developed  techniques   similar  to those of \cite{birrell-davies,parker-toms}, but we will use the later    method due to    Gosh \cite{gosh}  which is closer in spirit  to \cite{ZS} and appears to be  simpler and more efficient.

The time independent  ``in''  vacuum  state $\ket{0}$ is defined by  $\hat a_p\ket{0}= \hat b_p\ket{0}=0$, and, for any fixed    $p$ and  before the $A_x(t)$ field appears, the   corresponding  solutions  ${\bm \psi}_+(p,t)$, ${\bm \psi}_-(p,t)$ are those that have time dependence
  \be
{\bm  \psi}_+(p,t)= {\bm \psi}_+(p)e^{-i\epsilon t}, \quad {\bm \psi}_-(p,t)= {\bm \psi}_-(p)e^{+i\epsilon t}, \quad \epsilon= \sqrt{p^2+m^2}.
\ee
The initial $ {\bm \psi}_\pm (p)$  therefore coincide with the    positive and negative energy eigenvectors ${\bm \chi}_\pm(p)$    of the  hamiltonian matrix $H$ with $A_x=0$.  

In the course of the  time evolution caused by the field,  an occupied initially-negative-energy mode  ${\bm \psi}_-(p,t)={\bm \chi}_-(p) e^{i\epsilon t}$ will acquire a   positive energy  component and become
 \be
 {\bm \psi}_-(p,t) = \alpha(t) {\bm \chi}_-(t) e_+ + \beta(t) {\bm \chi}_+(t) e_-, 
\ee
where
\be
e_\pm(t)=  e^{\pm i \int_{-\infty}^t \epsilon(\tau) d\tau}.
\ee
The ${\bm \chi}_\pm(t)$ are most   conveniently chosen to be normalized positive and negative energy eigenvectors of the instantaneous  Hamiltonian matrix $H(t)$. The linear independence of ${\bm \chi}_\pm(t)$ then uniquely defines $\alpha(t)$ and $\beta(t)$.

The normalized instantaneous eigenvectors    are 
\be
 {\bm \chi}_+ = \frac 1{\sqrt{2\epsilon}} \left[\begin{matrix} \sqrt{\epsilon+\tilde p}\cr {\rm sgn}(m) \sqrt{\epsilon-\tilde p}\end{matrix}\right], \quad  {\bm \chi}_-=\frac 1{\sqrt{2\epsilon}} \left[\begin{matrix} \sqrt{\epsilon-\tilde p}  \cr - {\rm sgn}(m)\sqrt{\epsilon+\tilde p}\end{matrix}\right]
 \ee 
 where $\tilde p(t)= p-A_x(t)$. They obey 
 \be
 H(t)  {\bm \chi}_\pm = \pm \epsilon {\bm \chi}_\pm
 \ee
  with  $\epsilon = \sqrt{\tilde p^2+m^2}$, and  $|{\bm \chi}_\pm|^2=1$,   ${\bm \chi}^\dagger_+{\bm  \chi}_-={\bm \chi}^\dagger_-{\bm\chi}_+=0$. 
   The instantaneous eigenvectors do not obey the time evolution equation, but they still depend on $t$ as a parameter. Because they  are normalized and  have real entries they are  orthogonal to their derivative with respect to $t$. We can therefore most easily compute these  derivatives by using   the eigenstate perturbation formula   
\be
\dbrak{m}{\delta n}=\frac{\eval{m}{\delta H}{n}}{E_n-E_m}
\ee
to  find  the  projection of  $\dot {\bm \chi}_\pm$ on the  other eigenvector, and hence compute $\dot {\bm \chi}_\pm$ itself.
For example
\bea
\dbrak{\chi_+}{\dot \chi_-}&=& -\frac{\eval{\chi_+}{\dot H}{\chi_-}}{2\epsilon},\nonumber\\
&=&-\frac 1{2\epsilon}  \,{\bm \chi}_+^T 
 \left[ \begin{matrix} E(t) &0 \cr 0 &-E(t) \end{matrix}\right]  {\bm \chi}_- , \nonumber\\
&=&- \frac{|m| E(t) }{2\epsilon^2}.\
\eea  
The result is  
 \be 
 \frac {d}{dt} {\bm \chi}_+= +\left(\frac{E|m|}{2\epsilon^2}\right) {\bm \chi}_-, \quad \frac {d}{dt} {\bm \chi}_-= -\left(\frac{E|m|}{2\epsilon^2} \right){\bm \chi}_+.
 \ee
 Insert    
\be
{\bm \psi}(t)= \alpha(t) {\bm \chi}_- e_+ + \beta(t) {\bm \chi}_+ e_-
\ee
into the equation of motion (\ref{EQ:diracEOM}) to get   
\bea
0&=& \dot \alpha  {\bm \chi}_- e_+ + \alpha \dot {\bm \chi}_- e_+ + \dot \beta {\bm \chi}_+ e_- +  \beta \dot {\bm \chi}_+ e_-\nonumber\\
&=& \dot \alpha {\bm \chi}_- e_+- \alpha \left(\frac{E|m|}{2\epsilon^2} \right) {\bm \chi}_+e_+ + \dot \beta {\bm \chi}_+ e_- + \beta\left(\frac{E|m|}{2\epsilon^2}\right)  {\bm \chi}_-e_-.
\eea
From  the coefficients of the of the linearly independent ${\bm \chi}_\pm$ we read off that 
\be
\dot \alpha = -\left(\frac{E|m|}{2\epsilon^2}\right)\beta e_-^2, \quad \dot \beta  = +\left(\frac{E|m|}{2\epsilon^2}\right)\alpha e_+^2.
\ee
 Except for the relative minus sign  --- due to preserving $|\alpha|^2+|\beta|^2=1$ instead of $|\alpha|^2-|\beta|^2=1$ --- this is of the same form as the Bose case   (\ref{EQ:dot-alpha}).

When $\alpha \approx 1$ the  $\beta$ coefficient has a very rapid $\propto e_+^2$ phase evolution so that the time dependence of the adiabatically small  $\beta(t) {\bm \chi}_+ e_-$ term is close to that of the leading $\alpha(t) {\bm \chi}_- e_+$ term --- just as it is in the bosonic case.

The slowly varying quantities 
\be
\sigma = |\beta|^2, \quad \tau=i(\alpha \beta^* e_+^2 - \alpha^* \beta e_-^2), \quad \upsilon = (\alpha \beta^* e_+^2 + \alpha^* \beta e_-^2),
\label{EQ:S3-hopf}
\ee
now  label   points on the Bloch sphere
\be
(1-2\sigma)^2+ \tau^2+\upsilon^2=1
\ee
 and obey  the  spherical  Bloch equations  
\bea
\dot \sigma &=& F\upsilon,\nonumber\\
\dot \tau&=&- 2\epsilon \upsilon,\nonumber\\
\dot \upsilon&=& 2F(1-2\sigma) +2\epsilon \tau,
\label{EQ:spherical_bloch}
\eea
where $F= E|m|/2\epsilon^2= \dot {\tilde p}| m|/2\epsilon^2$, 
and in  the distant zero-field past  $\sigma=\tau=\upsilon=0$. 

Just as the bosonic  squeezed states are parameterized by points on a hyperboloid, the 2-mode fermionic squeezed state  of the form (\ref{EQ:2-mode-fermion}) correspond to  points on the Bloch sphere. The map (\ref{EQ:S3-hopf}) taking $\alpha,\beta$ to points on the sphere is now the classic Hopf map ${\rm SU}(2) \to S^3$. The equations (\ref{EQ:spherical_bloch}) can therefore be interpreted as ${\bm \psi}_-(p,t)$ being a 2-mode (positive and negative energy) squeezed version of the ground state of one particle Hamiltonian 
$H(t)$,  and $\ket{0}_{\rm in}$ being a squeezed version of the ground state of the corresponding instantaneous   many-body Hamiltonian.

As we are interested in the low energy response to the external field, we   will focus on the case in which the time evolution of the hamiltonian  is relatively slow compared to the mass gap.
 To obtain  an adiabatic expansion we again rearrange the Bloch equations (\ref{EQ:spherical_bloch}) as 
\bea
\sigma&=&  \int_0^t F[t'] \upsilon dt'\nonumber\\
\tau&=&- \frac{2F(1-2 \sigma) -\dot \upsilon}{2\epsilon}\nonumber\\
\upsilon&=&- \frac{\dot \tau}{2\epsilon},
\eea
and expand in inverse powers of $\epsilon(t)$ 
\bea
\sigma&=& \sigma_{\{2\}}+ \sigma_{\{4\}}+\ldots\nonumber\\
 \tau&=& \tau_{\{1\}}+ \tau_{\{3\}}+\ldots\nonumber\\
 \upsilon &=& \upsilon_{\{2\}}+ \upsilon_{\{4\}}+\ldots.
\eea
 
Regarding  $F$ as being $O[\epsilon^0]$   the    recursion relations are
\bea
\sigma_{\{n\}}&=& \int_0^t F[t']\upsilon_{\{n\}}dt',\nonumber\\
\upsilon_{\{n\}}&=&- \dot\tau_{\{n-1\}}/2\epsilon,\nonumber\\
\tau_{\{n+1\}}&=&\frac{4F \sigma_{\{n\}}+\dot \upsilon_{\{n\}}}{2\epsilon},\nonumber
\eea
with $\tau_{\{1\}}=-F/\epsilon$.

From the  matrix elements 
 \bea
{\bm \chi}^T_+ \sigma_1 {\bm \chi}_+=-{\bm \chi}^T_- \sigma_1 {\bm \chi}_-&=&  \frac{m}{\epsilon}\nonumber\\
{\bm \chi}^T_+ \sigma_2 {\bm \chi}_+= \phantom{-}{\bm \chi}^T_- \sigma_2 {\bm \chi}_-&=&  0\nonumber\\
{\bm \chi}^T_+ \sigma_3 {\bm \chi}_+=-{\bm \chi}^T_- \sigma_3 {\bm \chi}_-&=&\frac{\tilde p}{\epsilon}.
\eea
and 
\bea 
 {\bm \chi}^T_+ \sigma_1 {\bm \chi}_- = \phantom{-} {\bm \chi}^T_- \sigma_1 {\bm \chi}_+ &=& -\frac{\tilde p}{\epsilon},\nonumber\\
 {\bm \chi}^T_+ \sigma_2 {\bm \chi}_-=- {\bm \chi}^T_- \sigma_2 {\bm \chi}_+&=&i\, {\rm sgn}(m), \nonumber\\
 {\bm \chi}^T_+\sigma_3 {\bm \chi}_-= \phantom{-}{\bm \chi}^T_-\sigma_3 {\bm \chi}_+&=& \frac {|m|} \epsilon
 \eea
  we find, for each $p$-mode, 
\bea
{{\bm \psi}^\dagger \sigma_2 {\bm \psi}} &=& \alpha^* \beta e_-^2 {\bm \chi}_- \sigma_2 {\bm \chi}_+ +\alpha \beta^* e_+^2{\bm \chi}_+\sigma_2{\bm \chi}_-\nonumber\\
&=& -i\, {\rm sgn}(m) ( \alpha^* \beta e_-^2  -\alpha \beta^* e_+^2)\nonumber\\
&=& {\rm sgn} (m)\tau(t)\nonumber\\
&=&- \frac{mE}{2\epsilon^3}+\hbox{higher order}.
\eea

To compute the expectation values of the field operators we need to sum the contributions of the  filled negative-energy sea by integrating over $p$. For a time independent $E$ field we only need the lowest order contribution to $\tau$ and find that 
\bea
2m\eval{0}{\hat \psi^\dagger \sigma_2 \hat \psi}{0}&=&- \int_{-\infty}^{\infty} \frac{dp}{2\pi} \frac{m^2E}{\epsilon^3}\nonumber\\
&=&-E \int_{-\infty}^{\infty} \frac{dp}{2\pi} \frac{m^2}{\left((p+Et)^2+m^2\right)^{3/2}}\nonumber\\
&=&- \frac{E}{\pi }. 
\eea
Thus we find the anticipated result that
\be
2m \eval{0}{\hat \psi^\dagger \sigma_2\hat \psi}{0} + \frac 1{\pi}E=0, \quad \hbox{$E$  constant.}
\ee
 
Now consider a  time-dependent $E$ field.
Using the equations above we find that for any  $p$ mode  the matrix elements are given in terms of $\sigma$, $\tau$, $\upsilon$ as 
\bea
{\bm \psi}^\dagger \sigma_1 {\bm \psi} &=&(2\sigma-1) \frac {m}\epsilon - \upsilon \frac {\tilde p} \epsilon\nonumber\\
{\bm \psi}^\dagger \sigma_2 {\bm \psi} &=&{\rm sgn}(m) \tau.
\nonumber\\
{\bm \psi}^\dagger \sigma_3 {\bm \psi} &=&(2\sigma-1) \frac {\tilde p}\epsilon + \upsilon \frac {|m|} \epsilon
\eea
 So, using   
 \bea
 F&=&|m|\dot {\tilde p}/2\epsilon^2\nonumber\\
 \frac {d\epsilon}{dt}&=& \frac{\tilde p\dot{\tilde  p}}{\epsilon},\nonumber\\
 \frac{d}{dt}\frac 1 \epsilon &=& - \frac{\tilde p\dot {\tilde p}}{\epsilon^3},\nonumber\\
 \frac{d}{dt} \frac {\tilde p} \epsilon&=& \frac{\dot {\tilde p} m^2  }{\epsilon^3},\
 \eea
  and  the Bloch equations (\ref{EQ:spherical_bloch})  we find that the combination
 \be
 {\mathcal A}\stackrel{\rm def}{=} \partial_t({\bm \psi}^\dagger \sigma_3 {\bm \psi}) - 2m {\bm \psi}^\dagger \sigma_2 {\bm \psi},
 \ee
 which is expected to give the contribution of the $p$ mode to the anomaly, actually evaluates to 
 \bea
 {\mathcal A}&=& \partial_t\left((2\sigma-1) \frac{\tilde  p}\epsilon + \upsilon  \frac {|m|} \epsilon\right)- 2|m| \tau\nonumber\\
 &=& 2\dot \sigma \frac {\tilde p} \epsilon + (2\sigma-1)\frac{\dot {\tilde p} m^2 }{\epsilon^3} + \dot \upsilon  \frac {|m|} \epsilon - \upsilon {|m|} \frac{\dot {\tilde p} \tilde p}{\epsilon^3}-2|m|  \tau\nonumber\\
  &=&2F \upsilon \frac {\tilde p} \epsilon +(2\sigma-1)2F\frac {|m|} \epsilon  +(2F(1-2\sigma)+2\epsilon \tau) \frac {|m|}\epsilon -2F \upsilon \frac {\tilde p} \epsilon -2|m| \tau\nonumber\\ 
 &=&0.
 \eea
 That this expression is zero  is inevitable. The Bloch equations encode   the original field equations and  the c-number version of the chiral charge non-conservation equation  follows from these  field equations. Consequently,  when considered   {\it mode-by-mode\/}, the anomaly appears to be  zero. The  anomaly is non-zero, however, because the operations of integration over $p$ and taking the  time derivative do not necessarily  commute.
 
 For the rest of this section we will  assume that $m>0$, so $|m|=m$ and ${\rm sgn}(m)=1$, and  
again expand\footnote{We use the symbol $j$  for for the per-mode chiral-charge density and $J$ for the total  because  in 1+1 dimensions the chiral-charge {density} coincides with  the  particle-number {current}.}
 \be
 j\stackrel{\rm def}{=}{{\bm \psi}^\dagger \sigma_3 {\bm \psi}}
 \ee
 in  inverse powers of $\epsilon$ 
 \be
j = (2\sigma-1) \frac {\tilde p}\epsilon + \upsilon \frac {m} \epsilon= j_{\{1\}}+j_{\{3\}}+j_{\{5\}}+\ldots
 \ee
 As   $\sigma_{\{0\}}=\upsilon_{\{0\}}=0$, we have
 \be
 j_{\{1\}}= - \frac{\tilde p}{\epsilon}.
 \ee
For  $n\ge 2$, we have
 \be
 j_{\{n+1\}}(\tilde p,t) = 2\sigma_{\{n\}}(\tilde p,t) \left(\frac {\tilde p}\epsilon\right) + \upsilon_{\{n\}}(\tilde p,t) \left(\frac {m} \epsilon\right).
 \ee
 
 The corresponding contributions to the chiral charge are are given by summing over the occupied $p$-modes as
 \be
 J_{\{n+1\}}(t) \stackrel{\rm def}{=} \int_{-\infty}^\infty \frac{dp}{2\pi} j_{\{n+1\}} (\tilde p,t).
 \ee
 For example the recurrence relations give 
 \be
 j_{\{3\}}(\tilde p,t)=- \frac 58 \frac{m^2 \tilde p E^2}{ (m^2+\tilde p^2)^{7/2}} + \frac 1 4 \frac{m^2 \dot E}{(m^2+\tilde p^2)^{5/2}} 
   \ee
 The integral over $p$ is sufficiently convergent at  large $p$ that we can shift the integration variable $p\to \tilde p= p-A(t)$ without altering the value of the integral, and so find   
  \be
 J_{\{3\}}=\frac{\dot E}{6\pi m^2}.
 \ee
We can similarly shift the integration variable  and  integrate   
 \be
 \tau_{\{3\}}(\tilde p) =\frac{m \ddot E}{8 \left(m^2+\tilde p^2\right)^{5/2}}-\frac{5 m
   p E \dot E}{4 \left(m^2+\tilde p^2\right)^{7/2}}-\frac{3 m
   E^3}{8 \left(m^2+\tilde p^2\right)^{7/2}}+\frac{9 m \tilde p^2
   E^3}{4 \left(m^2+\tilde p^2\right)^{9/2}}+\frac{m^3
   E^3}{16 \left(m^2+\tilde p^2\right)^{9/2}}
\ee
 to find 
\be
T_{\{3\}}\stackrel{\rm def}{=}\int_{-\infty}^\infty \frac {dp}{2\pi}\tau_{\{3\}}(\tilde p)= \frac{\ddot E}{12 \pi  m^3}.
\ee
Consequently  $\partial_t J_{\{3\}}= 2m T_{\{3\}}$.

The same is true of 
\be
T_{\{5\}}=\int_{-\infty}^\infty \frac {dp}{2\pi}\tau_5(p)=\frac{4 E^2 \ddot E+8
   E {\dot E}^2-m^2 E^{(4)}}{60 \pi m^7}
   \ee
and 
\be
J_{\{5\}}=\frac{4 E^2 \dot E-m^2 E^{(3)}}{30 \pi m^6},
\ee
so   $\partial_t J_{\{5\}}=2m T_{\{5\}}$.  

The  interchange of  derivative and integral is indeed legitimate  for any $n>1$, 
but  an issue occurs for $j_{\{1\}}$ and $\tau_{\{1\}}$. As we have already seen 
\be
T_{\{1\}}= -\int_{-\infty}^\infty \frac {dp}{2\pi}\frac {E m}{2(p^2+m^2)^{3/2}}=-\frac E{2 \pi m}.
\ee
The problem is that  the integral for the expectation of the chiral charge 
\be
J_{\{1\}}\stackrel{\rm ?}{=}\int_{\infty}^\infty \frac {dp}{2\pi} j_1(\tilde p) =-\int_{-\infty}^\infty \frac {dp}{2\pi} \frac{\tilde p}{\sqrt{\tilde p^2+m^2}}
\ee
is convergent, but only {\it conditionally\/} convergent\footnote{The same is true in the  diagram calculation. The associated Feynman integral is linearly divergent by power counting, but becomes  conditionally convergent  after evaluating the gamma-matrix traces.}.  The value of the integral   depends on how we treat the large-$p$ limits,  and is  therefore   ambiguous. If we elect to remove the ambiguity by choosing limits that are symmetric about $p=0$  
\bea
 \int_{-\infty}^\infty \frac {dp}{2\pi} j_1(\tilde p)&\stackrel{\rm def}{=}&  \lim_{\Lambda\to \infty} \left\{\int_{-\Lambda}^\Lambda \frac {dp}{2\pi}j_1(\tilde p)\right\}=\lim_{\Lambda\to \infty} \left\{-\int_{-\Lambda}^\Lambda \frac {dp}{2\pi} \frac{\tilde p}{\sqrt{\tilde p^2+m^2}}\right\}\nonumber\\
 \eea 
 then, for  $A(t)=0$,   
 \bea
 \lim_{\Lambda\to \infty} \left\{-\int_{-\Lambda}^\Lambda \frac {dp}{2\pi} \frac{p}{\sqrt{p^2+m^2}}\right\}
 \eea
 is zero by the $p\leftrightarrow -p$ symmetry.  The $A(t)\ne0$ integral, however,   requires  shifting the integration variable  $p\to \tilde p= p-A$ and leads to 
 \be
 \lim_{\Lambda\to \infty} \left\{-\int_{-\Lambda}^\Lambda  \frac {dp}{2\pi} \frac{(p-A)}{\sqrt{(p-A)^2+m^2}}\right\}= \lim_{\Lambda\to \infty} \left\{-\int_{-\Lambda-A}^{\Lambda-A} \frac {dp'}{2\pi} \frac{p'}{\sqrt{p'^2+m^2}}\right\}=\frac{A}\pi.
  \ee
 If we accept this  symmetric-cutoff definition  of $J_{\{1\}}$   we end up with  the anomaly-free $\partial_tJ_{\{1\}}= 2m T_{\{1\}}$ --- but the  arbitrariness is unsatisfying. More concerning is that    a time-independent shift in $A_x(t)$ is    a gauge transformation. If  we insist on preserving gauge invariance  we must   ensure that  the physical value of $J_{\{1\}}$  is unaffected  when $A_x$ is augmented by a constant. 
We can arrange this   gauge invariance by re-defining  the physical  chiral charge operator  to be  
\be
\hat J_{\rm phys}\stackrel{\rm def}{=}\hat \psi^\dagger \sigma_3 \hat \psi - \frac{A_x}{\pi}.
\ee
This new definition  must  be used even when $A_x$ becomes time dependent, so making $J_{\{1\}}\equiv 0$ and 
giving the anomalous    $0=\partial_t J_{\{1\}}= 2m T_{\{1\}} + E /\pi $. 

The necessary $-A_x/\pi$  c-number subtraction in the definition of the physical current is the source of the difference between the mode-by-mode  chiral-charge evolution equation and the anomalous  equation for the mode-summed current. It is also  this subtraction that allows the simple physical interpretation of the energy levels that cross zero in the $m=0$ case as being newly created particles  and holes --- even though the occupation numbers of the Heisenberg states, labeled by their initial ``$p$'' and counted by $\hat \psi^\dagger \sigma_3 \hat \psi$, remain  unchanged.  

Understanding that the  $E/\pi$  comes only from the lowest order of the adiabatic expansion reveals  why  the total charge created by the anomaly is  insensitive to how rapidly the  time evolution occurs.  

\subsection{Two space dimensions: the parity anomaly}

The interesting physics in 2+1 dimensions is the ``parity anomaly'' \cite{parity-anomaly} which arises from the fact that in odd spacetime dimensions a Dirac mass term  violates space-inversion symmetry.    The result is a current at right angles to the applied electric field  with  the direction of the current  depending  on the sign of the mass term. This effect is usually derived by computing a one-loop triangle diagram  and extracting from it a Chern-Simons effective action \cite{parity-anomaly2,odd-anomaly}. We can, however, also obtain it  by a small modification of the results from the previous section.

Keeping  the electric field  parallel to the $x$ axis the 2+1 dimensional  Dirac hamiltonian is 
\bea
H(t)&=& \sigma_3 (p_x-A_x)  + p_y \sigma_2 + m\sigma_1 \nonumber\\
&=& \left[\begin{matrix}p_x-A_x(t) & m-ip_y\cr m+ip_y & -(p_x-A_x(t))\end{matrix}\right].
 \eea
We have   
\bea
&=& \sigma_3 (p_x-A_x)  +M \sigma_1e^{i\sigma_3 \phi} \nonumber\\
&=& \sigma_3 (p_x-A_x)  +M (\sigma_1\cos \phi + \sigma_2 \sin\phi)\nonumber\\
 &=& e^{-i\sigma_3 \phi/2}(\sigma_3 (p_x-A_x)  + M\sigma_1) e^{i\sigma_3 \phi/2}
 \eea
 so if we define $M =\sqrt{p_y^2+m^2}$ and an  angle $\phi$ by setting $M \cos\phi = m$,  $M\sin\phi =p_y$, 
 we find that the solutions to
 \be
 i\partial_t {\bm \psi}= H(p_x, p_y,t)  {\bm \psi}
 \ee
 are simply 
 \be
  {\bm \psi} = e^{-i\sigma_3 \phi/2} {\bm \psi}_{1d}
  \ee
  where $ {\bm \psi}_{1d}$ are the solutions for the one-dimensional case, but  with the $m$ appearing there replaced by $M =\sqrt{p_y^2+m^2}$.
  
  For fixed $p_x, p_y$ we have 
  \be
  j_x= {\bm \psi}^\dagger \sigma_3 {\bm \psi}=  {\bm \psi}_{1d}^\dagger \sigma_3 {\bm \psi}_{1d}
  \ee
  but the more interesting transverse component $j_y$ is given by 
  \bea
  j_y&=& {\bm \psi}^\dagger{\bm \sigma}_2{\bm \psi}= {\bm \psi}_{1d}^\dagger \left(e^{i\sigma_3 \theta/2}{\bm \sigma}_2e^{-i\sigma_3 \theta/2}\right){\bm \psi}_{1d}\nonumber\\
  &=&\cos\phi \,{\bm \psi}_{1d}^\dagger{\bm \sigma}_2{\bm \psi}_{1d}+ \sin\phi\, {\bm \psi}_{1d}^\dagger{\bm \sigma}_1{\bm \psi}_{1d}\nonumber\\
  &=& \frac{m}{\sqrt{p_y^2+m^2}}{\bm \psi}_{1d}^\dagger{\bm \sigma}_2{\bm \psi}_{1d}+ \frac{p_y}{\sqrt{p_y^2+m^2}}{\bm \psi}_{1d}^\dagger{\bm \sigma}_1{\bm \psi}_{1d}\nonumber\\
  &=& \frac{m} {\sqrt{p_y^2+m^2}} {\rm sgn}(M) \tau(t) + \frac{p_y}  {\sqrt{p_y^2+m^2}} \left((2\sigma(t)-1)\frac{\sqrt{p_y^2+m^2}}{\sqrt{\tilde p_x^2+p_y^2 +m^2}} - \upsilon(t) \frac{\tilde p_x}{\sqrt{\tilde p^2_x +p^2_y +m^2}}\right).\nonumber\\
  \eea
  
To compute the parity anomaly current we simply  substitute our results from one dimension  and perform the $p_x$, $p_y$ integrals. The $p_y$ integral of the last term ${\bm \psi}_{1d}^\dagger{\bm \sigma}_1{\bm \psi}_{1d}$ is zero, so we only need the $\tau$ term
\be
 \frac{m}{\sqrt{p_y^2+m^2}}\tau.
\ee

In particular, at lowest order, we have 
\be
j_y^{\{1\}}(p_x,p_y) = \frac{m}{\sqrt{p_y^2+m^2}} \tau_1= \frac{m}{M} \left( -\frac{E_xM}{2(p^2+m^2)^{3/2}} \right)
\ee
so
\bea
J_y^{\{1\}} &=& \int \frac{d^2p}{(2\pi)^2} j_y^{\{1\}}(p_x,p_y),\nonumber\\
&=& - \frac {m E_x}{4\pi} \int_0^\infty \frac{p\,dp}{(p^2+m^2)^{3/2}},\nonumber\\
&=& -\frac {m E_x}{4\pi |m|},\nonumber\\
&=& - {\rm sgn}(m) \frac{E_x}{4\pi},
\label{EQ:parity-current}
\eea
which is the correct coefficient.

Higher terms in the adiabatic expansion are all total derivatives
\bea
J_y^{\{3\}} &=& {\rm sgn}( m)  \frac{\partial}{\partial t} \left( \frac 1{48 \pi m^2}  \dot E_x\right),\nonumber\\
J_y^{\{5\}} &=& {\rm sgn}( m)  \frac{\partial}{\partial t} \left( - \frac{1}{320\pi m^4} E_x^{(3)} +\frac{1}{96 \pi m^6} E_x^2 \dot E_x\right),\nonumber\\
J_y^{\{7\}} &=& {\rm sgn}( m)  \frac{\partial}{\partial t} \left( \frac{1}{1792 \pi m^6} E_x^{(5)} -
\frac{1}{768 \pi m^8}(5 \dot E_x^3 +20 E_x \dot E_x \ddot E_x +5 E_x^2 E_x^{(3)}) +\frac{3}{128 \pi m^{10}} E_x^4 \dot E_x \right),\nonumber\\
\eea
and so on.

The parity anomaly provides  the  simplest  illustration of the 
general Callan-Harvey anomaly-inflow mechanism   \cite{callan-harvey}. 
 In 2+1 dimensions a domain wall across which the fermion mass changes sign traps  a 1+1 dimensional  chiral fermion which possesses  a charge-conservation anomaly 
 \be
 \partial_t \rho_{\rm wall}= \frac{E_{\parallel}}{2\pi},
 \ee
  that is one-half of the $E/\pi$ Dirac-particle chiral anomaly.  Here $E_{\parallel}$ is the component of the two dimensional electric field  parallel to the domain wall. 
  The  charge  $\rho_{\rm wall}$ is  not appearing from nowhere, but is supplied by  twice (because the wall has two sides)  the   Hall-effect-like  parity-anomaly  current (\ref{EQ:parity-current}) which is perpendicular to the wall. The change in sign of the fermion mass across the wall means that the  currents  from the two sides are in opposite directions and therefore add.  If the  wall is parallel to the $x$ axis, and if  $E_x$ is zero at $t=\pm \infty$  then {\it total\/}  
charge per unit length that appears on the domain wall is given by 
\be
\Delta \rho_{\rm wall}^0= \frac{1}{2\pi} \int_{-\infty}^\infty  E_x(t) dt.
\ee
This expression for the accumulated charge  is exact because  the total derivatives in the higher order contributions to $J_y$ make no net  contribution -- everything comes from the leading order term only.

\subsection{Three  dimensional vector currents and vacuum polarization}
\label{SEC:three-dimensions}

 In 3+1 dimensions, and with the external electric field parallel to the $z$ axis, it turns out to  convenient to change the  spinor basis so that the 4-by-4  Dirac Hamiltonian  
\be
H(t)= \alpha^1 p_1+\alpha^2 p_2+\alpha^3 (p_3-A_z(t))  +\beta m
\ee
becomes 
\be
H(t)= \left[\begin{matrix}p_3-A_z(t) & M\cr M^\dagger & -(p_3-A_z(t))\end{matrix}\right], \quad M=M^\dagger = p_1 \sigma_1 +p_2 \sigma_2 +m\sigma_3.
\ee

The $\beta\equiv \gamma^0$ matrix must   be the coefficient of $m$ and so
 \be
 \gamma^0=  \left[\begin{matrix} &\sigma_3  \cr \sigma_3 &\end{matrix}\right]= \sigma_1 \otimes \sigma_3.
 \ee
 Similarly
 \bea
 \alpha^1&\equiv &\gamma^0 \gamma^1=  \sigma_1\otimes \sigma_1 \Rightarrow \gamma^1= {\mathbb I}\otimes \sigma_3 \sigma_1\nonumber\\
 \alpha^2&\equiv &\gamma^0 \gamma^2 = \sigma_1\otimes \sigma_2 \Rightarrow \gamma^2= {\mathbb I}\otimes \sigma_3 \sigma_2\nonumber\\
 \alpha^3 &\equiv &\gamma^0 \gamma^3 = \sigma_3\otimes {\mathbb I}\,\,\, \Rightarrow \gamma^3= \sigma_1\sigma_3\otimes \sigma_3.
 \eea
 We see that distinct  $\gamma$'s mutually anticommute, and 
 \be
 (\gamma^0)^2=1, \quad  (\gamma^1)^2=(\gamma^2)^2=(\gamma^3)^2=-1,
 \ee
 so we have a valid, if non-standard, representation of the Dirac   algebra.
 
 The  chiral symmetry operation should rotate $m$ and so must be   
 \bea
&& \left[\begin{matrix} e^{-i\sigma_3 \theta}& \cr & e^{i\sigma_3 \theta}\end{matrix}\right]\left[\begin{matrix}p_3-A_z(t) &p_1 \sigma_1 +p_2 \sigma_2 +m\sigma_3 \cr p_1 \sigma_1 +p_2 \sigma_2 +m\sigma_3 & -(p_3-A_z(t))\end{matrix}\right] \left[\begin{matrix} e^{i\sigma_3 \theta}& \cr & e^{-i\sigma_3 \theta}\end{matrix}\right]\nonumber\\
 &&\qquad\qquad = \left[\begin{matrix}p_3-A_z(t) &p_1 \sigma_1 +p_2 \sigma_2 +(me^{-2i\theta})\sigma_3 \cr p_1 \sigma_1 +p_2 \sigma_2 +(me^{2i\theta})\sigma_3 & -(p_3-A_z(t))\end{matrix}\right].
 \eea
 Thus
 \be
  \left[\begin{matrix} e^{i\sigma_3 \theta}& \cr & e^{-i\sigma_3 \theta}\end{matrix}\right]\to e^{i\theta \gamma^5}
  \ee
 and  
 \be
 \gamma^5 =  \left[\begin{matrix} \sigma_3 & \cr & -\sigma_3\end{matrix}\right]= \sigma_3\otimes \sigma_3.
 \ee
 We see that the $\gamma^\mu $ and $\gamma^5$ anticommute as they should.

The $p$-mode contribution to the expectation of the vector current is  then
 \be
 j_z = {\bm \psi}^\dagger (\sigma_3\otimes {\mathbb I}) {\bm \psi},
 \ee
 and 
the chiral charge density and chiral current density are respectively 
\bea
j^5_0 &=&  {\bm \psi}^\dagger \gamma^5 {\bm \psi}={\bm \psi}^\dagger ( \sigma_3\otimes \sigma_3) {\bm \psi},\nonumber\\
j^5_3 &=&  {\bm \psi}^\dagger \gamma^0\gamma^3\gamma^5  {\bm \psi}= {\bm \psi}^\dagger (\sigma_3 \otimes \mathbb I)(\sigma_3\otimes \sigma_3) {\bm \psi}= {\bm \psi}^\dagger ({\mathbb I}\otimes \sigma_3) {\bm \psi}.
\label{EQ:3d-chiral-current}
 \eea
 Similarly, the  bilinear that appears on the right-hand-side of the chiral charge (non)-conservation law is 
 \be
i  \bar{\bm \psi} \gamma^5 {\bm \psi}=i{\bm \psi}^\dagger \gamma^0\gamma^5 {\bm \psi}=i {\bm \psi}^\dagger (\sigma_1 \otimes \sigma_3 )( \sigma_3\otimes \sigma_3 ) {\bm \psi}= {\bm \psi}^\dagger ( \sigma_2\otimes {\mathbb I}) {\bm \psi} .
\label{EQ:3d-chiral-bilinear}
 \ee

 The advantage of the unconventional basis choice is that the  $t$-parameterized snapshot eigenfunctions of  $H(t)$ factorize. They are  a tensor product of our previous 2-spinor eigenstates with   a second orthonormal set of   2-spinor eigenstates   ${\bm \lambda}_\alpha $, $\alpha=\pm$,  of the two-by-two matrix $M$. These  have  eigenvalues $\mu =\pm q$ where 
 \be
 q=\sqrt{m^2+p_1^2+p^2_2} =\sqrt{m^2+p_\perp^2}.
\ee 
We find that  
\bea
 {\bm \chi}_+(t) &=& \frac 1{\sqrt{2\epsilon}} \left[\begin{matrix} \phantom{{\rm sgn}(\mu)}\sqrt{\epsilon+\tilde p}\,{\bm  \lambda}_\alpha \cr {\rm sgn}(\mu) \sqrt{\epsilon-\tilde p} \,{\bm  \lambda}_\alpha \end{matrix}\right] = \frac 1{\sqrt{2\epsilon}} \left[\begin{matrix} \phantom{{\rm sgn}(\mu)}\sqrt{\epsilon+\tilde p}\cr {\rm sgn}(\mu) \sqrt{\epsilon-\tilde p}\end{matrix}\right] \otimes {\bm  \lambda}_\alpha\nonumber\\ 
  {\bm \chi}_-(t)&=&\frac 1{\sqrt{2\epsilon}} \left[\begin{matrix} \phantom{-{\rm sgn}(\mu)}\sqrt{\epsilon-\tilde p}\, {\bm \lambda}_\alpha  \cr - {\rm sgn}(\mu)\sqrt{\epsilon+\tilde p}\,{\bm \lambda}_\alpha\end{matrix}\right]=\frac 1{\sqrt{2\epsilon}} \left[\begin{matrix} \phantom{-{\rm sgn}(\mu)}\sqrt{\epsilon-\tilde p}  \cr - {\rm sgn}(\mu)\sqrt{\epsilon+\tilde p}\end{matrix}\right]\otimes { \bm \lambda}_\alpha.\nonumber\\
 \eea
    Here again   $\tilde p= p_3-A(t)$, but now $\epsilon = \sqrt{\tilde p_3(t)^2 +\mu^2}$.
    
 Again we expand 
 \be
 {\bm \psi}(t)= \alpha(t) {\bm \chi}_-(t) e_+ +\beta(t) {\bm \chi}_+ e_-.
 \ee  
 The evolution equations for $\alpha$, $\beta$, $\sigma$, $\tau$, $\upsilon$ are the same as in 1+1 dimensions  except that $\epsilon$ is the 3+1 energy, and $|m|$ must be replaced by $|\mu|$. 
 Also the  the vector-current expectations   need to be multiplied by $2$ to take into account both ${\bm \lambda}_+$ and ${\bm \lambda}_-$.
 
The vector current density  is   
 \bea
 j_z &=& {\bm \psi}^\dagger (\sigma_3\otimes {\mathbb I}) {\bm \psi},\nonumber\\
 &=& 2 \left((2\sigma-1) \frac {\tilde p}\epsilon + \upsilon \frac {|\mu| } \epsilon\right).
 \eea
We find that
\be
j^{\{3\}}_z= 2\left(-\frac 58 \frac{\mu^2 p_3 E^2}{(\mu^2+ p_3^2)^{7/2}}+ \frac 14 \frac{\mu^2 \dot E}{(\mu^2+p_3^2)^{5/2}}\right).
\ee
Because of the $p_3$ in its numerator, the first term cancels in the momentum  integration. The second   contributes   
\bea
J^{\{3\}}_z&=& \int \frac{d^3 p}{(2\pi)^3} j^{(3)} \nonumber\\
&=& \dot E \cdot \frac 12  \int \frac{d^3 p}{(2\pi)^3} \frac{p_1^2+p_2^2+m^2}{(p_1^2+p_2^2+p_3^2+m^2)^{5/2}}\nonumber\\
&=& \dot E\cdot \frac 12\left( \frac 2 3 \int \frac{d^3 p}{(2\pi)^3} \frac{p_1^2+p_2^2+p_3^2}{(p_1^2+p_2^2+p_3^2+m^2)^{5/2}} +  \int \frac{d^3 p}{(2\pi)^3} \frac{m^2}{(p_1^2+p_2^2+p_3^2+m^2)^{5/2}}\right)\nonumber\\
&=& \dot E\cdot \frac 1{4\pi^2}\left(\frac 23 \int_0^\infty \frac{p^4 \,dp}{(p^2+m^2)^{5/2}}+    \int_0^\infty \frac{m^2 p^2 \,dp}{(p^2+m^2)^{5/2}}\right)
\label{EQ:unregulated}
\eea
The second integral in the parentheses in the last line of (\ref{EQ:unregulated}) is independent of $m$ and equal to $1/3$. Scaling $p\to mp$ suggests that the   first term 
is also  independent of $m$ ---  but as the integral  is   logarithmically divergent this is an unsafe conclusion. To get a meaningful result we should   regulate $J^{\{3\}}_z$  by introducing a suitable high energy cutoff.  The simplest Lorentz- and gauge-invariant  way to do this is to follow Pauli-Villars  and  subtract from  the integrand in (\ref{EQ:unregulated}) the same expression  but with  $m^2$  everywhere replaced by $\Lambda^2$. Then
\be
\int_0^\infty \frac{p^4 \,dp}{(p^2+m^2)^{5/2}}\to I\stackrel{\rm def}{=}\int_0^\infty \left( \frac{p^4}{(p^2+m^2)^{5/2}}-  \frac{p^4}{(p^2+\Lambda^2)^{5/2}}\right)dp.
\ee
The new integrand  falls off as $1/p^2$ at large $p$ so the integral  is  convergent and   easily evaluated by Frullani's method 
\bea 
I&=&\lim_{M\to \infty} \left\{\int_0^M \left( \frac{p^4}{(p^2+m^2)^{5/2}}-  \frac{p^4}{(p^2+\Lambda^2)^{5/2}}\right)dp\right\}\nonumber\\
&=& \lim_{M\to \infty} \left\{\int_0^{M/m}  \frac{\rho^4\, d\rho }{(\rho^2+1)^{5/2}}-  \int_0^{M/\Lambda }\frac{\rho^4 \,d\rho }{(\rho^2+1)^{5/2}}\right\}\nonumber\\
&=& \lim_{M\to \infty}\left\{\int_{M/\Lambda}^{M/m}  \frac{\rho^4\, dp }{(\rho^2+1)^{5/2}}\right\}\nonumber\\
&=&  \lim_{M\to \infty}\left\{\int_{ 1/\Lambda}^{1/m}  \frac{\rho^4\, d\rho }{(\rho^2+M^{-2})^{5/2}}\right\}\nonumber\\
&=& \int_{ 1/\Lambda}^{1/m}  \frac{d\rho }{\rho} \nonumber\\
&=& \ln  \left(\Lambda/m\right).
\eea
The   resulting cut-off-dependent  current  
\be
{\bf J}(\Lambda)=  =  \frac{\dot {\bf E}}{8\pi^2}\left( \frac 23     \ln \left(\frac{\Lambda^2}{m^2}\right)\right)
\label{EQ:divergent}
 \ee
can be interpreted as ${\bf J}(\Lambda)= \dot {\bf P}(\Lambda)$ where 
 \be
{\bf P}(\Lambda)=\frac{ {\bf E}}{8\pi^2}\left( \frac 23     \ln \left(\frac{\Lambda^2}{m^2}\right)\right)
\ee
 is the   polarization of the vacuum.    
 The associated cut-off-dependent  dielectric constant 
 is  the source of  charge renormalization  (see appendix \ref{SEC:effective} for details).
 
 Higher order terms  in the adiabatic series are all finite --- but more complicated. For example  
\bea
j^{\{5\}}_z({\bf p})&=&2\left(-\frac{\mu ^2 E^{(3)}}{16 \left(\mu
   ^2+p_3^2\right)^{7/2}}+\frac{21 \mu ^2 p_3
   \dot E^2}{32 \left(\mu
   ^2+p_3^2\right)^{9/2}}+\frac{7 \mu ^2 p_3
  E  \ddot E }{8 \left(\mu
   ^2+p_3^2\right)^{9/2}}\right.\nonumber\\&&\left.
   +\frac{19 \mu ^2 E^2
   \dot E}{16 \left(\mu
   ^2+p_3^2\right)^{9/2}}-\frac{117 \mu ^2 p_3^2
   E^2 \dot E }{16 \left(\mu
   ^2+p_3^2\right)^{11/2}}
   -\frac{3 \mu ^4 E^2
   \dot E}{32 \left(\mu ^2+p_3^2\right)^{11/2}}\right.\nonumber\\
   &&\left.
   -\frac{57 \mu ^2 p_3
   E^4}{16 \left(\mu
   ^2+p_3^2\right)^{11/2}}+\frac{303 \mu ^2 p_3^3
   E^4}{32 \left(\mu
   ^2+p_3^2\right)^{13/2}}+\frac{57 \mu ^4 p_3
   E^4}{128 \left(\mu
   ^2+p_3^2\right)^{13/2}}\right).\nonumber\\
  \eea
 The expression for  $j_z^{\{7\}}({\bf p})$ contains  69 terms and $j_z^{\{9\}}({\bf p})$ nearly 5,000, but the resulting expressions for the current 
 \be
 J_z^{\{n\}}= \int \frac{d^3 p}{(2\pi)^3} j^{\{n\}}(\bf p)
 \ee 
 are relative compact:
 \bea
 J_z^{\{5\}}&=&\frac{1}{2\pi^2}\left(\frac{1}{15} \frac{E^2 \dot E}{ m^4} - \frac{1}{30} \frac{E^{(3)}}{ m^2}\right),\nonumber\\
 J_z^{\{7\}}&=&\frac{1}{2\pi^2}\left( \frac{2}{21} \frac{E^4 \dot E}{m^8}- \frac{2}{63} \frac{(\dot E)^3}{m^6}
 -\frac{8}{63}\frac{E \dot E \ddot E}{m^6}-\frac{2}{63}\frac{E^2 E^{(3)}}{ m^6} + \frac{1}{280}\frac{E^{(5)}}{m^4}\right),\nonumber\\
 J_z^{\{9\}}&=&\frac 1{2\pi^2}\left(-\frac{E^{(7)}}{1890
   m^6}+\frac{E^{(5)} E^2}{84 
   m^8}-\frac{8 E^{(3)}E^4}{75 
   m^{10}}+\frac{16  E^6 \dot E}{45 
   m^{12}}-\frac{16 E^2 (\dot E)^3}{25 
   m^{10}}+\frac{E^{(4)} E \dot E}{14
    m^8}\right.\nonumber\\
   &&\left.+\frac{5 E^{(3)} E
   \ddot E}{42  m^8}+\frac{2 E^{(3)}
   (\dot E)^2}{21  m^8}-\frac{64 E^3
  \dot E \ddot E}{75  m^{10}}+\frac{11
  \dot E (\ddot E)^2}{84  m^8}\right).
\label{EQ:uptoninth}
 \eea
 The expressions in Eq. ({\ref{EQ:uptoninth})   contain contributions to the current that can be verified by comparison with other methods.  
 For example, if we retain only  the terms linear in $E$ and its  derivatives $E^{(n)} \equiv d^nE /dt^n$ we find
\bea
j_z^{\{5\}}({\bf p})&=& 2\left(- \frac 1{16} \frac{\mu^2 E^{(3)}}{(\mu^2+p_3^2)^{7/2}}+O(E^2)\right),\nonumber\\
j_z^{\{7\}}({\bf p})&=& 2\left(+\frac 1{64} \frac{\mu^2 E^{(5)}}{(\mu^2+p_3^2)^{9/2}}+O(E^2)\right),\nonumber\\
j_z^{\{9\}}({\bf p})&=& 2\left(- \frac{1}{256} \frac{\mu^2 E^{(7)}}{(\mu^2+p_3^2)^{11/2}}+O(E^2)\right),\nonumber
\eea
where the pattern of  rational-number coefficients  is clear.
In the  resulting currents
\bea
J_z^{\{5\}}&=&  \int \frac{d^3 p}{(2\pi)^3} j_{z}^{\{5\}} =- \frac 1 {30} \frac{1}{2\pi^2} E^{(3)} +O(E^2),\nonumber\\
J_z^{\{7\}}&=&  \int \frac{d^3 p}{(2\pi)^3} j_{z}^{\{7\}} = + \frac{1}{280} \frac{1}{2\pi^2} E^{(5)}+O(E^2),\nonumber\\
J_z^{\{9\}}&=&   \int \frac{d^3 p}{(2\pi)^3} j_{z}^{\{9\}} =-\frac{1}{1890} \frac{1}{2\pi^2} E^{(7)}+O(E^2),
\label{EQ:2-leg}
\eea
the sequence of numerical  fractions   is more obscure, but, as we will show in in appendix \ref{SEC:effective},  they are the coefficients occurring  in  the the series expansion of the  two-point  vacuum polarization diagram in powers of the frequency.

Another set of terms that can be confirmed by other   methods are those containing  an arbitrary power of $E$, one power of $\dot E$, and  no higher derivatives of $E$. 

For example the coefficient of $E^2 \dot E$ in $j^{\{5\}}({\bf p})$ is 
\bea
{\rm coef}(E^2 \dot E)^{\{5\}}&=&2 \left(\frac{19 \mu ^2 }{16 \left(\mu
   ^2+p_3^2\right)^{9/2}}-\frac{117 \mu ^2 p_3^2
    }{16 \left(\mu
   ^2+p_3^2\right)^{11/2}}
   -\frac{3 \mu ^4 }{32 \left(\mu ^2+p_3^2\right)^{11/2}}\right)\nonumber\\
   &=& 2\left(
   +\frac{19 (p^2 \sin^2\theta +m^2) }{16 \left(p^2+m^2\right)^{9/2}}-\frac{117( p^2 \sin^2\theta +m^2) p^2 \cos^2\theta
    }{16 \left(p^2 
   +m^2\right)^{11/2}}
   -\frac{3 (p^2 \sin^2\theta +m^2)^2 }{32 \left(p^2+m^2\right)^{11/2}}\right).\nonumber\\
    \eea
    where $\theta$ is the angle between ${\bf p}$ and the $z$-axis. The integration measure becomes 
    \be
    \frac{d^3p}{(2\pi)^3}= \frac {1}{(2\pi)^2} \sin \theta d\theta p^2 dp
    \ee
 and gives  
 \be
 J^{\{5\}}_z= {\dot E} \frac{1} {2 \pi^2} \frac {1}{15}\frac{E^2}{ m^4}+ \ldots.
 \label{EQ:schwinger1}
 \ee
 The analogous calculation with $ j^{\{7\}}$ gives a term
 \be 
J^{\{7\}}_z={\dot E}\frac{1}{2\pi^2} \frac {2}{21} \frac{E^4}{ m^8}+ \ldots.
\label{EQ:schwinger2}
\ee
Both these terms  agree   with those in the current calculated from the Euler-Heisenberg-Schwinger one-loop effective action (see appendix \ref{SEC:effective}).

\subsection{Three-dimensional chiral  currents and their anomaly}
\label{SEC:3d-anomaly}

In three space dimensions the uniform electric field version of the anomalous chiral current (non)-conservation is expected be
\be
\partial_t\, {\hat \psi}^\dagger (\sigma_3\otimes \sigma_3) {\hat \psi}= 2m\, \hat{\psi}^\dagger (\sigma_2\otimes {\mathbb I}) {\hat \psi}+ \frac 1{2\pi^2} {\bf E}\cdot {\bf B}.
\ee
In the absence of a magnetic field, the   vacuum expectation of this equation is 
satisfied by all three terms being zero. The expectation of the current on the LHS is zero because, from (\ref{EQ:3d-chiral-current}), each $p_3$ mode  has a factor of  
\be
{\bm \lambda^\dagger_+}\sigma_3  {\bm \lambda_+}+ {\bm \lambda^\dagger_-}\sigma_3  {\bm \lambda_-}
\ee
where ${\bm \lambda_\pm}$ are the  normalized eigenvectors of 
\be
M= \sigma_1 p_1+\sigma_2 p_2+ \sigma_3 m
\ee
with eigenvalue $\mu= \pm \sqrt{|p_\perp|^2+m^2}$.
This  sum of expectations  is zero because sandwiching the identity  
\be
\sigma_3 M + M\sigma_3 =2m{\mathbb I}
\ee
between the $\pm$ eigenvectors and subtracting  gives 
\be
|\mu| ( {\bm \lambda^\dagger_+}\sigma_3  {\bm \lambda_+}+ {\bm \lambda^\dagger_-}\sigma_3  {\bm \lambda_-})=0.
\ee
 The expectation of the RHS is zero because, from   (\ref{EQ:3d-chiral-bilinear}) , 
\be
{\bm \psi}^\dagger(\sigma_2\otimes {\mathbb I})  {\bm \psi} \propto {\rm sgn}(\mu) 
\ee
 and we must sum over both signs of the eigenvalue $\mu$ when we include the ${\bm \lambda_\pm}$ factors.

 To get a non-trivial result we must include a magnetic field.  If we take  the   vector potential to be 
 \be
{\bf A}(x,y)= (By/2,-Bx/2,0)
\ee
then  ${\bf B}= (0,0,- B)$ and  a  classical positively charged particle will orbit anticlockwise about  the $z$ direction.  It is well known \cite{rabi,landau-dirac}  that in such a field the spectrum   of  the operator $M$  is highly degenerate,  being  an ensemble of Landau levels  in which each level  eigenvalue $\mu$  has degeneracy $|{\bf B}|/2\pi$ per unit area perpendicular to the magnetic field.
One set of eigenfunctions of $M$ is  
\be 
{\bm \lambda}_{l,\pm}\propto \left[\begin{matrix} (\mu+|m|) r^{-1} e^{-i\theta}\cr iB\end{matrix}\right] r^l e^{il\theta} \exp\left\{ - \frac{Br^2}{4}\right\} , \quad l>0
\ee
with eigenvalues $\mu= \pm \sqrt{2l B+m^2}$.  These eigenfunctions  correspond to the positively  charged  particles circling   the $z$ axis at a distance  $\vev{r}= \sqrt{2l/B}$. The other eigenstates in the same Landau level can be thought of as as describing orbits of  the same radius but with different centers.  There are no $l<0$ modes because they would be  singular at the origin.

The features   that lead  all terms in the chiral current to cancel still hold, so the $l>0$ eigenmodes make no contribution to the anomaly equation. 
The $l=0$ case is special, however. The $\mu= -|m|$ eigenvector   
\be
{\bm \lambda}_0=   \frac{1}{\sqrt{2\pi}} \left[\begin{matrix} 0\cr 1\end{matrix}\right] \exp\left\{ - \frac{Br^2}{4}\right\}
\ee
is finite  at $r=0$ but 
the $\mu=+|m|$   is not  because of the $r^{-1}$ in the upper component. As only one of the $\mu=\pm |m|$ pair is allowed, there is no cancellation in the currents and we recover exactly the  situation explored in sections \ref{SEC:1-d-chiral}  and 
\ref{SEC:Two-level-adiabatic} --- except that the RHS of the one-dimensional anomaly equation (\ref{EQ:1-d-chiral-anomaly}) we must replace 
\be
\frac{E_z}{ \pi} \to \frac{{\bf E}\cdot {\bf B}}{2\pi^2}
\ee   
to take into account the $B/2\pi$ per-unit-area Landau-level degeneracy.  Note that the fermions still have a mass-gap given by $m$ and so the same interplay between time derivative and $2m \vev{\hat\psi^\dagger \sigma_2\hat\psi}$ occurs as in \ref{SEC:Two-level-adiabatic}.

\section{Conclusion} We have shown that a simple quantum mechanics adiabatic expansion can be used to capture  the non-trivial physics of the chiral and parity anomalies for massive fermions. Although similar in spirit to the well known spectral flow interpretation of the chiral anomaly, the presence of a fermion mass both thwarts the flow and at the same time controls  the accuracy of the expansion, so  allowing us to see that in both cases the anomaly arises only from the lowest  term in the expansion.

\section{Acknowledgements} This work was not directly supported by any funding agency, but it would not have been possible without resources provided by the Department of Physics at the University of Illinois at Urbana-Champaign.

\appendix
\appendixpage

\section{The driven oscillator} 
\label{SEC:driven}
For completeness we include a brief discussion of the well-known coherent state solution of the  constant-frequency  harmonic oscillator driven by a time-dependent  external force.  The hamiltonian can be written in terms of $\hat a$, $\hat a^\dagger$ as 
\be
H(t)= \Omega \hat a^\dagger \hat a +F(t) \hat a^\dagger +F^*(t) \hat a.
\label{EQ:driven-oscillator-hamiltonian}
\ee
 The associated unitary evolution operator  $U(t)={\mathcal T} \!\exp\{-i \int_0^t H(t) dt\}$  obeys the equation
\be
i\partial_t U = \{\Omega \hat a^\dagger \hat a +F(t) \hat a^\dagger +F^*(t) \hat a\}U
\ee
and we may seek a solution in the form
\bea
U&=& e^{i\theta} e^{-{\textstyle \frac 12}|z|^2}e^{z\hat a^\dagger} e^{-z^*\hat a} e^{-i\omega \hat a^\dagger \hat a},\nonumber\\
U^{-1} &=& e^{-i\theta} e^{+{\textstyle \frac 12}|z|^2}e^{i\omega \hat a^\dagger \hat a} e^{z^*\hat a} e^{-z\hat a^\dagger}. 
\eea 
Then, using
\bea
e^{-i\varphi \hat a^\dagger \hat a}\left[\begin{matrix}\hat a\cr \hat a^\dagger\end{matrix}\right] e^{i\varphi \hat a^\dagger \hat a}&=&  \left[\begin{matrix} \hat a e^{+i\varphi}\cr\hat  a^\dagger  e^{-i\varphi}\end{matrix}\right],\nonumber\\
e^{-\lambda \hat a^\dagger} \,\hat a\, e^{+\lambda \hat a^\dagger}&=& \hat a+\lambda,\nonumber\\
 e^{+\lambda^* \hat  a} \hat a^\dagger e^{-\lambda^* \hat a}&=& \hat a^\dagger+\lambda^*,
\eea
we find 
 \be
iU^{-1}\partial_t U= i e^{i\omega \hat a^\dagger \hat a}\left[ \frac 12 (-\dot z z^*-  \dot z^* z) +\dot z (\hat a^\dagger+z^*)- \dot z^* \hat a-i\dot \omega \hat a^\dagger \hat a+i\dot \theta\right] e^{-i\omega \hat a^\dagger \hat a},
\label{EQ:coherent-EOM}
\ee
and  
\be
U^{-1} \{\Omega \hat a^\dagger \hat a +F(t) \hat a^\dagger +F^*(t) \hat a\}U= e^{i\omega \hat a^\dagger \hat a}[\Omega(\hat a^\dagger +z^*)(\hat a+z)+F(t)(\hat a^\dagger+z^*)+F^*(t)(\hat a+z)]e^{-i\omega \hat a^\dagger \hat a}.
\ee
Comparing coefficients of $\hat a^\dagger \hat a$, $\hat a^\dagger$, $\hat a$ and $1$, we read off 
that \cite{wei-norman}
\bea
\dot \omega &=& \Omega,\nonumber\\
i\dot z&=&\Omega z+F,\nonumber\\
-i\dot z^* &=&\Omega z^* +F^*,\nonumber\\
\dot \theta &=&{\textstyle\frac i2} (\dot z z^*- \dot z^* z)-( \Omega z^* z+ Fz^*+F^*z).
\eea
The accumulating phase  
\be 
 \theta(t) =\int_0^t\left\{ \frac{i}{2} (\dot z z^*-\dot z^*z)-(\Omega z^* z +F z^*+F^*z)\right\}dt
\ee
 considered as a functional $\theta = S[z(t)]$ of the path $z(t)$  is the classical  action  whose variation gives the equations of motion for $z$ and $z^*$ in (\ref{EQ:coherent-EOM}).  Thus   $S[z(t)]$ for the actual trajectory  is Hamilton's principal function solution of the Hamilton-Jacobi equation \cite{van-hove,stone-matsumoto}. 
 
The unitary displacement operator  
\be
D(z)\stackrel{\rm def}{=} e^{z\hat a ^\dagger - z \hat a}= e^{-{\textstyle \frac12}|z|^2} e^{z\hat a^\dagger}e^{-z^*\hat a} 
\ee
acts on the ground state to create a conventional  (unsqueezed) coherent state  
\be 
\ket {z}=  e^{z\hat a^\dagger -z^* \hat a}\ket{0} =e^{-{\textstyle \frac 12}|z|^2}e^{z\hat a^\dagger} \ket{0}
\ee
that   obeys   $\hat a\ket{z}= z\ket z$.
Under our  time evolution  
\be
\ket{0}\mapsto U[t]\ket{0}=e^{i\theta(t)} \ket{z(t)}, 
\ee
so  if we start in the ground state $\ket{0}$ the  external driving force will always leave the system  in   a  coherent state. As  
\be
\hat a = \frac 1{\sqrt 2}\left( \sqrt\Omega\,  \hat x +\frac{i}{\sqrt\Omega} \hat p\right),
\ee
and the coherent state maps $\hat a\mapsto z$,
the real and imaginary parts of the  complex parameter $z(t)$ track   the classical motion in $x$, $p$ phase space. 

Note that allowing  the parameter $\Omega$ appearing in  (\ref{EQ:driven-oscillator-hamiltonian})  to become time dependent does {\it not\/} have the same effect as the time dependence of  the  frequency $\Omega(t)$  appearing in   the  original oscillator  Hamiltonian  (\ref{EQ:harmonic-oscillator}). We need  to include additional $a^2$ and $({a^\dagger})^2$ terms in  (\ref{EQ:driven-oscillator-hamiltonian}) to change the oscillator frequency   because 
changing the frequency    while  keeping the operators $\hat p$ and $\hat x$ fixed redefines $\hat a$ and $\hat a^\dagger$.

\section{Dirac effective action and renormalization  in 1+3 dimensions}
\label{SEC:effective} 

Throughout the main text we have focussed on non-interacting fermions and scaled the $A_\mu$ field so as to  set the particle charge to unity.  We do this because in a    general non-abelian gauge theory the coefficient of $A_\mu$  becomes  the Lie-algebra  matrix generator corresponding to the group representation in which the fermion  lives ---  and for  the ${\rm U}(1)$ group of electromagnetism the representations are labelled by integers.   
This rescaling has the effect that in the QED path integral
\be
Z= \int d[A]d[\bar \psi]d[\psi] \exp\{iS[A]\}
\ee
the pure-gauge contribution to the action  becomes 
 \be 
  S_{\rm Maxwell}[A]
  =-\frac{1}{4 e_0^2} \int d^4x F_{\mu\nu}F^{\mu\nu} = \frac{1}{2e_0^2}\int d^4 x \left({\bf E}^2-{\bf B}^2\right).
\ee
The (bare) coupling constant $e^2_0$ now   appears in its natural location where it governs the magnitude of  gauge-field fluctuations.  The integration over $\bar\psi $, $\psi$ adds to   the 
pure-gauge action  the  fermionic {effective action\/}    
\be
S_F= - i \ln {\rm Det}({\fsl D}[A]+m). 
\ee
where ${\fsl D}= i\gamma^\mu (\partial_\mu+iA_\mu)$, $A_\mu = (\phi, -{\bf A})$.

One special case in which $S_F$ can be evaluated in closed form  is when the field strength $F_{\mu\nu}= \partial_\mu A_\nu-\partial_\nu A_\mu$ is constant  \cite{euler-schwinger,dunne-euler}. 
For a constant electric field we have  
\be
S_F[E]= VT \,{\mathcal L}_F[E]
\ee
where $V$ and $T$ are the volume and total time, respectively, and the effective Lagrangian density is 
\bea
{\mathcal L}_{\rm F}[E]&=& 
-\frac 1{8\pi^2} \int_0^\infty    \frac{ds}{s^3} \{sE\cot(sE)\}e^{-m^2s}\nonumber\\
&=& \frac{1}{8\pi^2} \left(\frac 13 {E^2} \ln \left(\frac {\Lambda^2}{m^2}\right) +\frac{1}{45} \frac{E^4}{m^4}+ \frac{4}{315} \frac{E^6}{m^8}+\frac{8}{315} \frac{E^8}{m^{12}}\ldots\right). 
\label{EQ:schwinger}
\eea
In obtaining the second line we  have expanded 
\be
(sE)\cot(sE ) = 1-\frac{(sE)^2}{3}-\frac{(sE)^4}{45}-\frac{2 (sE)^6}{945}- \frac{(sE)^8}{4725}+ \ldots,
\ee
and dropped the infinite negative  Dirac sea  contribution from the first term. The second term has then been regulated\footnote{Strictly we need {\it three\/} Pauli-Villars massive particles to gauge-invariantly regulate both divergent terms.} using the Frullani integral
\be
\int_0^\infty \frac {e^{-m^2 s}- e^{-\Lambda^2 s}} {s} ds = \ln \left(\frac {\Lambda^2}{m^2}\right).
\ee
The remaining terms are all finite.

The cutoff dependent term is proportional to ${\bf E}^2$, and this would become $ {\bf E}^2-{\bf B}^2 $ had we included a constant magnetic field. We can therefore combine  the cut-off dependent  term with the free-field Maxwell action by   replacing   
\be
\frac{1}{2 e_0^2}\int ({\bf E}^2- {\bf B}^2) d^d x \to \frac{1}{2 e_R^2} \int ({\bf E}^2- {\bf B}^2) d^d x
\ee
where the renormalized coupling constant $e^2_R$ is defined  by
\be
\frac{1}{ e_R^2}=  \frac{1}{ e_0^2}+ \frac{1}{8\pi^2} \frac 23 \ln \left(\frac {\Lambda^2}{m^2}\right).
\label{EQ:relation}
\ee
This relation can be written as 
\be
e^2_R = \frac{e^2_0}{1+ \frac{e^2_0}{8\pi^2} \frac 23\ln  \left(\frac {\Lambda^2}{m^2}\right)},\quad
{e^2_0}= \frac{e^2_R}{1- \frac{e^2_R}{8\pi^2} \frac 23\ln  \left(\frac {\Lambda^2}{m^2}\right)}.
 \ee
In terms of the fine-structure constant $\alpha \equiv  e_R^2/4\pi$  and its  bare value $\alpha_0 = e_0^2/4\pi$  Eq.\ (\ref{EQ:relation}) becomes   
\be 
\alpha= \frac {\alpha_0}{1+ \frac {\alpha_0 }{3\pi}\ln \left(\frac {\Lambda^2}{m^2}\right)},
\ee
which gives us the  one-loop beta function for QED
\be
\beta_{\rm \tiny one-loop}(\alpha)\stackrel{\rm def}{=} \left(\frac{\partial  \alpha }{\partial \ln m}\right)_{\Lambda, \alpha_0}= \frac{2 }{3}\frac{\alpha^2}{\pi}.
\ee

 Although (\ref{EQ:schwinger}) was derived under the assumption that $E$ is constant, we can  
use $J_z=  \delta {S_F}/{\delta A_z}$ and 
\be
\delta \int  E^n d^4x =  \int  n E^{n-1} (-\partial_t \delta A_z) d^4x=  n(n-1) \int   (E^{n-2} \dot E)\delta A_z d^4x
\ee
to compute the current $\propto \dot E$ induced by a slowly  varying electric field. The reason is  that  higher-order $\dot E$ corrections to (\ref{EQ:schwinger})   give rise in $J_z$ to terms proportional to $\ddot E$ or $(\dot E)^2$ or higher. Exploiting  this observation we   find from (\ref{EQ:schwinger}) the following contribution to the induced current
\be
J_z = \frac{\dot E}{2 \pi^2} \left( \frac 2 {12} \ln \left(\frac {\Lambda^2}{m^2}\right)+ \frac {1}{15}\frac{ E^2 }{ m^4} +  \frac{2}{21} \frac{ E^4}{ m^8} +\frac{16}{45} \frac{ E^6}{ m^{12}}\ldots\right)
\label{EQ:J_z}
\ee
which agrees with the corresponding terms  in  the adiabatic expansion in Eqs.\ 
(\ref{EQ:schwinger1}) and (\ref{EQ:schwinger2})

A second  special case is the two-point diagram with arbitrary space-time dependent $A_\mu({\bf x}, t) $ fields
\bea
S_{{\rm F}2}&\equiv&  \frac 12 \int d^4x d^4 x' A_\mu(x) \Pi_{\mu\nu}(x,x') A_\nu(x').
\eea 
In momentum space  the kernel $\Pi_{\mu\nu}$  is given by \cite{itzykson-zuber}  
\be
\Pi_{\mu\nu}(k^2)= (g_{\mu\nu}k^2- k_\mu k_\nu) \Omega(k^2,m, \Lambda),
\ee
where $\Lambda$ is a momentum-space cut-off and 
\be
\Omega(k^2,m, \Lambda)=
 - \frac{1}{2\pi^2}\int_0^1 dx\, x(1-x) \left\{- \ln \frac{\Lambda^2}{m^2}+\ln\left( 1- x(1-x) \frac{k^2}{m^2}\right)\right\}.
\ee
For a time-dependent electric field $\propto e^{i\omega t}$ we have $k_\mu=( \omega,0,0,0)$  and  $g_{zz}k^2=-\omega^2$, and so in frequency space
\be 
{\mathcal L}_{{\rm F}2}= \frac 1{4\pi^2} A_z(\omega)A_z(-\omega) \int_0^1 dx\, x(1-x) \left\{- \ln \frac{\Lambda^2}{m^2}+\ln\left( 1- x(1-x) \frac{\omega^2}{m^2}\right)\right\}
\ee
Recalling that  $E_z=-\partial_t A_z$ we see that this gives rise to a current
\be
J_z(\omega) = \dot E_z(\omega)\frac{1}{2\pi^2}\int_0^1 dx\, x(1-x) \left\{ \ln \frac{\Lambda^2}{m^2}-\ln\left( 1- x(1-x) \frac{\omega^2}{m^2}\right)\right\}.
\ee
As  
\be
\int_0^1 dx\, x(1-x)= \frac 1 6
\ee
we have   
the   cut-off dependent    contribution to the vacuum polarization current  
\be
J_{z, {\rm \tiny divergent}}= \dot {\bf P}=\frac {\dot E} {8\pi^2} \left(\frac 23  \ln\left( \frac{\Lambda^2}{m^2}\right)+\ldots \right) 
\ee
that we obtained  in Eq.\ (\ref{EQ:divergent}) and (\ref{EQ:J_z}).

The non-divergent parts of ${\mathcal L}_{{\rm F}2}[E]$ come from  
\bea
I\left(\frac{\omega^2}{m^2}\right)&=&- \int_0^1 dx\, x(1-x) \ln \left(1- x(1-x) \frac{\omega^2}{m^2}\right)\nonumber\\
&=&\sum_{n=1}^\infty \left(\frac{\omega^2}{m^2}\right)^n \frac 1 n \int_0^1dx\,  x^{n+1}(1-x)^{n+1}\nonumber\\
&=&\sum_{n=1}^\infty \left(\frac{\omega^2}{m^2}\right)^n \frac 1 n \frac{[(n+1)!]^2}{(2n+3)!}\nonumber\\
&=& \frac{1}{30} \left(\frac{\omega^2}{m^2}\right)+ \frac 1{280} \left(\frac{\omega^2}{m^2}\right)^2+ \frac{1}{1890} \left(\frac{\omega^2}{m^2}\right)^3+\ldots
\eea
After  one takes into account that each factor of $\omega^2$ corresponds to a $-d^2/dt^2$, the sequence of rational coefficients agree with those found in Eq.\ (\ref{EQ:2-leg}). The resulting series has radius of convergence $\omega^2 = 4 m^2$, which is determined by a  branch-point in the function  $\Pi_{\mu\nu}(\omega^2, m^2)$.   The  discontinuity across the branch cut for $\omega^2 >  4 m^2$ gives the cross-section  for  particle-hole  pair  creation {\it via\/}  parametric resonance.

\section{Vacuum polarization  and  energy density}

In Euclidean signature  a time-independent effective-action density has a physical interpretation as the  vacuum-energy density. In Minkowski signature the action density becomes a  Lagrangian   density whose interpretation in terms of  energy  is not so clear.
It is interesting, therefore, to relate  the constant-field effective Lagrangian density  ${\mathcal L}_F[E]$  in (\ref{EQ:schwinger}) to the vacuum  energy expressed as  a function of the polarization.

 Recall that 
in a  Minkowski-signature  action functional  the coupling of the gauge field to the  current  is given by $\int  {\bf A}\cdot {\bf J}\,d^d x $.  
Therefore, 
if  $S_F   = \int {\mathcal L}_Fd^dx $  then
the space component of the current is given by  the functional derivative
\be
J_z({\bf x},t)=  \frac{\delta S_F}{\delta A_z({\bf x},t)}.
\ee
For spatially constant $E$ 
\be
\delta S_F[E]= \int d^dx {\mathcal L}_F'[E] (-\partial_t \delta A_z(t))=  \int d^dx {\mathcal L}_F''[E] \dot E  \delta A_z, 
\ee
where the prime denotes a derivative with respect to $E$.
The corresponding current density is therefore 
\be
{\bf J}=  {\mathcal L}_F''(E)\dot {\bf E}.
\label{EQ:slow-current}
\ee
As the current is related to the vacuum polarization by ${\bf J}=\dot {\bf P}$, Equation  (\ref{EQ:slow-current})
 implies that the vacuum polarization is related to the action-density ${\mathcal L}_F[E]$ by $P[E]={\mathcal L}_F'[E]$.

The associated vacuum energy-density  ${\mathcal E}$ as function of $P$  is found from the work required for an external $E$ field to adiabatically  create the   given polarization density over  a  period of time $\tau$. This  work is 
\bea
{\mathcal E} &=& \int_0^\tau J E dt \nonumber\\
&=& \int_0^\tau {\mathcal L}_F''(E)E \dot E dt\nonumber\\
& =& \int_0^E   {\mathcal L}_F''(\varepsilon)\varepsilon  d\varepsilon\nonumber\\
& =& \int_0^E (- {\mathcal L}_F'[\varepsilon])d\varepsilon +\left[\varepsilon{\mathcal L}_F'[\varepsilon]\right]_0^{E} \nonumber\\
&=& E {\mathcal L}_F'[E] -{\mathcal L}_F[E].
\label{EQ:legendre}
\eea
In the last line we assumed that  that ${\mathcal L}_F$ is zero when $E=0$.
The result (\ref{EQ:legendre})  is a  Legendre  transformation, so ${\mathcal E} \equiv  E{\mathcal L}_F'-{\mathcal L}_F$  is naturally  expressed in terms of  $P= {\mathcal L}_F'[E]$ as ${\mathcal E}[P]$.  We may then recover the electric field from the derivative 
\bea
\frac{d{\mathcal E}}{dP}&=& \frac {dE}{dP} P + E - \frac {d{\mathcal L}_F}{dP} \nonumber\\
&=& \frac {dE}{dP} P +E - \frac {dE}{dP}  \frac{d{\mathcal L}_F }{dE}\nonumber\\
&=&  \frac {dE}{dP} P +E - \frac {dE}{dP}  P\nonumber\\
&=& E.
\eea
These   transformations are just a slightly disguised version of the standard   relation between  the Lagrangian and  Hamiltonian. 
 
 The stability of the vacuum requires that   ${\mathcal E}[P]$ be a convex function of $P$ and this is property assured by the signs of the terms in (\ref{EQ:schwinger}) and that the Legendre transform of a convex function is convex.

\end{document}